\documentclass[a4paper,11pt]{article}
\usepackage{jheppub}
\usepackage{lineno}

\usepackage[utf8]{inputenc}
\usepackage{lipsum}
\usepackage{xcolor}
\usepackage{amsthm,amsfonts,mathrsfs}

\usepackage{tabularx}
\usepackage{booktabs}
\usepackage[capitalise]{cleveref} 
\usepackage{comment}
\usepackage{subfigure}

\usepackage[normalem]{ulem}
\usepackage{enumitem}
\usepackage{dsfont}

\definecolor{darkgreen}{rgb}{0.0,0.4,0.0}
\renewcommand{\l}{\left}
\renewcommand{\r}{\right}

\newcommand{\be}{\begin{equation}}
\newcommand{\ee}{\end{equation}}

\newcommand{\into}{\rightarrow}

\def\Amplitude{\mathcal{A}}

\begin{document}

\title{\textbf{Eikonal, nonlocality and regular black holes}}

\author{Mariano Cadoni,}
\emailAdd{mariano.cadoni@ca.infn.it}

\author{Lorenzo Herres,}
\emailAdd{lorenzo.herres@ca.infn.it}

\author{Leonardo Modesto,}
\emailAdd{leonardo.modesto@unica.it}

\author{Lorenzo Orlando,}
\emailAdd{lorenzo.orlando@ca.infn.it}

\author{\\ Mirko Pitzalis}
\emailAdd{mirko.pitzalis@ca.infn.it}

\affiliation{Dipartimento di Fisica, Universit\`a di Cagliari, Cittadella Universitaria, 09042 Monserrato, Italy}
\affiliation{I.N.F.N, Sezione di Cagliari, Cittadella Universitaria, 09042 Monserrato, Italy}

\abstract{ We investigate the leading gravitational eikonal in nonlocal $D$ dimensional theories of gravity. We analyze the simplest cases of $2\rightarrow2$ massless and massive scalar scattering at tree level, studying the effects of nonlocal form factors in the gravitational sector. We give an interpretation of our results in terms of geodesic motion in effective generalized Aichelburg-Sexl geometries for the massless case, and in smeared linearized Schwarzschild metrics for the massive case in the probe limit. Combining our results for the geometries at linearized level with general requirements about the behaviour of the solutions in the core,  we propose a nonlinear completion of the geometries. The resulting spacetimes describe  singularity-free, asymptotically flat deformations of the Schwarzschild solution with a de Sitter core. We also analyze the main geometric and thermodynamic features of these solutions.}
\maketitle

\section{Introduction}
In the late eighties and early nineties several investigations \cite{Amati:1987wq, Amati:1988tn,Amati:1993tb, Verlinde_1992,tHooft:1987vrq, Muzinich:1987in} addressed the problem of transplanckian-energy gravitational collisions, envisaged as Gedankenexperimente to probe interesting regimes and generic features in quantum gravity. While the dream was, and to a certain extent still remains, to gain control over the small impact parameter limit which is expected to lead classically to black hole formation, the main progress was instead achieved in the large (with respect to typical Schwarzschild radii $R_s$ of the process) impact parameter regime, which can be analyzed and smoothly connected to classical physics. The relevant high-energy, semiclassical limit, well captured in a relativistic setting by the \emph{eikonal} approximation, in recent years has also proven useful for studying collisions of astrophysically heavy compact objects, both in GR and in other theoretical laboratories \cite{DiVecchia:2020ymx,KoemansCollado:2019ggb,Parra-Martinez:2020dzs,AccettulliHuber:2020oou,DiVecchia:2021bdo}. This is an active line of research which is ever in need of new complementary approaches and computational techniques, especially in view of the rise of multi-messenger astrophysics that started with the first detection of gravitational waves \cite{Marion:2017enj,LIGOScientific:2017vwq}.\\
In this general setting, it may be of interest to extend the analysis to nonlocal theories of gravity \cite{Biswas:2011ar, Modesto:2011kw, Calcagni:2014vxa, BasiBeneito:2022wux}, which may  provide a well-behaved UV completion of general relativity.
The actions describing these theories come equipped with form factors which effectively resum an infinite number of higher derivative couplings and can be chosen so as to avoid new poles, ghostly or physical, and to tame the problematic UV behavior of Einstein gravity. Such form factors are usually taken, in momentum space, to be entire functions $e^{H(z)}$ of the dimensionless combination $z=\alpha p^2$, with $\alpha \equiv \ell^2$ a fixed characteristic length scale squared encoding the onset of nonlocal physics, and engineered to recover familiar local field theories in the $\alpha \rightarrow0$ limit. By treating the form factors and the nonlocality scale as tunable input data of the theory, however, we can imagine the latter to be even decoupled from the Planck scale $\ell_P$, giving a chance for their effects to survive the classical limit and show up, at intermediate impact parameters, in the usual gravitational IR observables, such as time delays or deflection angles. This is naturally similar in spirit to the seminal works of Amati, Ciafaloni and Veneziano on gravitational scattering in string theory \cite{Amati:1987uf,Amati:1988tn,Amati:1987wq,Amati:1992zb,Amati:1993tb}, in which $\alpha' \gg \ell_P^2$ is assumed   and the parameters  encoding classical (Post-Minkowskian) and stringy corrections are $\l(R_s/b\r)^{(D-2)} $ and $\alpha'/b^2$, respectively.\\
The primary tool at our disposal will be the gravitational \emph{eikonal} phase. The main idea (see \cite{DiVecchia:2023frv} for an excellent review) is that at high energies and fixed momentum transfer --- the Regge limit --- the S-matrix for a scattering  process is well approximated by an infinite sum to all orders in the loop expansion of leading terms in ladder and crossed ladder diagrams,  representing the exchange of an increasing number of almost on-shell soft gravitons \footnote{In string theory, actually the whole leading Regge trajectory gets exchanged at high energies.}. In impact parameter space, these contributions generically have the right combinatorics to be combined in a single phase, the leading eikonal phase, which is extracted from the Born amplitude and it is extremely large in the classical limit. This has two important effects: it unitarizes the elastic process, otherwise inconsistent with partial wave unitarity, and justifies a saddle point analysis from which observables like the classical impulse, deflection angles or time delays may be reliably extracted, up to classical, Post-Minkowskian corrections buried in subleading (in energy) terms of loop diagrams.\\
A captivating feature of the eikonal approach is the possibility of seeing, order by order, an effective curved geometry being reconstructed dynamically from computations carried out perturbatively around flat space. For example, the leading eikonal phase for massless scattering in ordinary GR is intimately connected to the phase shift of one of the two scatterers in the background of the emergent Aichelburg-Sexl \cite{Aichelburg:1970dh} shockwave produced by the other. An even cleaner instance is elastic scattering in the probe limit, where, roughly, one of the two objects is much heavier than the other. Here, backreaction of the probe can be safely neglected and one expects the eikonal to faithfully reproduce the physics of geodesic motion of the probe in the background produced by the heavy particle. For neutral, spinless objects in  ordinary Einstein gravity minimally coupled to matter, the expected geometry is Schwarzschild-Tangherlini, as one finds directly from the equations of motion of the theory. In the context of nonlocal gravitational theories, however, the field equations linking source and geometry seem to be prohibitively complicated and typically only linearized \cite{Biswas:2011ar,Giacchini_2019,Boos:2018bxf,Buoninfante:2018stt,Giacchini:2018gxp,Kolar:2020bpo} or approximate \cite{Modesto:2010uh,Modesto:2011kw,Buoninfante:2019swn,Bambi:2013gva} solutions can be found. The geometric approach is then not readily available, and agreement between the two methods, easily checked in the linearized approximation, can become a consistency condition to be imposed on putative exact solutions rather than derived from first principles.\\
Another widespread expectation is that nonlocality, as a mechanism, should be powerful enough to make all singularities in the geometry disappear, acting in this respect as a UV regulator \cite{Koshelev:2024wfk,Buoninfante:2018xiw,Edholm:2016hbt,Koshelev:2018hpt,Burzilla:2020utr,Giacchini:2016xns,Boos:2018bxf,Boos:2020qgg,Boos:2021suz}. It is thus reasonable to assume that regular black holes should play an important role in the complete exact solutions. This gives a quite natural point of contact with the rich literature on nonsingular black holes (see e.g. \cite{Carballo-Rubio:2025fnc,Cadoni:2023lqe,Cadoni:2023lum,Cadoni:2024rri,Cadoni:2026ejk,bardeen,Dymnikova:1992ux,Hayward:2005gi,Ayon-Beato:2000mjt,Bronnikov:2000vy,Carballo-Rubio:2018pmi,Carballo-Rubio:2022kad,Lan:2023cvz,Bambi:2023try,Bambi:2016wmo,Burzilla:2020bkx,Giacchini:2021pmr,Mo:2022szw,Zhou:2022yio,Zhou:2023lwc,dePaulaNetto:2023cjw,Burzilla:2023xdd,dePaulaNetto:2021axj,Bueno:2024dgm,Bueno:2025zaj,Fernandes:2025eoc,Fernandes:2025mic,Eichhorn:2025pgy,Bueno:2026dln}).
In fact, regular black holes may provide a consistent framework for solving conceptual and fundamental problems of gravitational physics, ranging from the singularity problem to the information paradox \cite{Cadoni:2023nrm,Cadoni:2023tse,Cadoni:2022chn,Akil:2022coa,Bonanno:2024wvb}.\\
In this paper we investigate the leading eikonal approximation in nonlocal, $D$-dimensional,  theories of gravity. Our goal is twofold. On the one hand we want to set up the general formalism that allows us to extract  the leading eikonal --- in the simplest case of massless and massive scalar scattering --- in the presence  of nonlocal  form factors in the gravitational sector. On the other hand we aim to reconstruct from results at the linearized level,  the  full nonlinear form of the spacetime geometry. An effort in this direction is made in the probe limit of the massive case  by using a combination of consistency conditions and suitable physical arguments  about the UV effects of nonlocality to constrain the possible form of the nonlinear exact completion of the geometry.
\\
The plan of the paper is as follows: in Section \ref{Massless scalar scattering} we present the models under scrutiny, we extract the leading eikonal for the massless case and discuss it mainly focusing, as we do for most of the paper, on the example of the Gaussian form factor, which is expected to capture essential features while still allowing analytic results. We highlight similarities and crucial differences with the string  case. We analyze the  S-matrix, discuss the fate of the usual 't Hooft poles found in massless scattering in four dimensional Einstein gravity, and make contact with motion in generalized Aichelburg-Sexl geometries.\\
In Section \ref{Massive scalar scattering} we generalize our analysis to the massive case, and quickly specialize to the probe limit, where we derive general formulae for our benchmark observable, the leading deflection angle. 
In Section \ref{sect:nonlinear} we describe our method to reconstruct the full exact spacetime geometry from the linearized solution of the massive probe limit. We  derive the form  of the metric and show that it represents a deformation of the Schwarzschild solution describing an asymptotically flat nonsingular black hole with a de Sitter  core. We also briefly discuss some geometric and thermodynamical features of our solutions. 
In Section \ref{sect:infrared} we analyze a class of non-analytic, infrared form factors, expected to capture very different physics. We present our conclusions and future directions in Section \ref{Conclusion}. In the appendices we fix conventions and kinematics (Appendix \ref{app:A}), check the exponentiation in the case of the simplest massless scattering (Appendix  \ref{app:B}), and we compute the leading deflection angle of spinless probes in specific deformations of the Schwarzschild black hole (Appendix  \ref{app:C}).

\section{Massless scalar scattering}\label{Massless scalar scattering}

As a starting point we consider the family of nonlocal gravity actions quadratic in the Ricci scalar $R$ and in the Ricci tensor $R_{\mu\nu}$, whose nonlocality is parametrized  by a single function $\omega$ \footnote{Here and in the following we shall omit the usual gauge-fixing terms in the action.}  
\begin{align}\label{starting_action}
    S_{g} &= \frac{1}{2\kappa_D^2}\int d^Dx\sqrt{|g|}\Big( R-\dfrac{1}{2}R\omega(-\alpha\Box)R+R_{\mu\nu}\omega(-\alpha\Box)R^{\mu\nu}\Big) \\
    \nonumber &= \frac{1}{2\kappa_D^2}\int d^Dx\sqrt{|g|}\Big( R+G_{\mu\nu}\omega(-\alpha\Box)R^{\mu\nu}\Big)
\end{align}
with $\kappa_D^2 =8\pi G_D$,  where $G_{D}$ is the $D$ dimensional Newton constant and  $\alpha \equiv \ell^2$ is the length scale squared characterizing nonlocality. In the second line we have also written the action in terms of the Einstein tensor $G_{\mu\nu}$. In principle, we could choose different functions for the last two terms in the first line of \eqref{starting_action}, but a common, simplifying  choice is as above. By parametrizing the function $\omega$ as follows
\begin{equation}\label{gammaBox}
    \omega(-\alpha\Box)=\dfrac{e^{H(-\alpha\Box)}-1}{\Box},
\end{equation}
it can be shown that, upon expanding the metric tensor around flat spacetime,
\begin{equation}
    g_{\mu\nu} = \eta_{\mu\nu} +2\kappa_D h_{\mu\nu},
\end{equation}
the graviton propagator behaves (in de Donder gauge and with the usual Feynman $i0$ prescription) as
\begin{equation}\label{grav_prop}
    G_{\mu\nu,\rho\sigma}(q^2) = -i\frac{e^{-H(\alpha q^2)}}{q^2-i0}P_{\mu\nu, \rho\sigma},
\end{equation}
with the familiar tensor structure
\begin{equation}
    P_{\mu\nu, \rho\sigma} = \frac{1}{2}\l(\eta_{\mu\rho}\eta_{\nu\sigma} + \eta_{\mu\sigma}\eta_{\nu\rho} - \frac{2}{D-2}\eta_{\mu\nu}\eta_{\rho\sigma} \r).
\end{equation}
The expression for the graviton propagator given by \cref{grav_prop} allows us to  interpret physically the  function $\omega$ parametrizing the nonlocality in terms of a nonlocal spreading of the graviton propagator.
Regarding entire (also called weakly nonlocal) form factors, common choices of $\omega(-\alpha\Box)$ contain analytic functions $H(-\alpha\Box)$ in \eqref{gammaBox} of the class \cite{Modesto:2011kw}
    \begin{align}\label{ff}
    H(\alpha q^2)=c\int_0^{p(\alpha q^2)} dy\,\dfrac{1-\xi(y)}{y}, \quad \xi(y)=\exp({-y^n}),
\end{align}
where $p(\alpha q^2)$ is a polynomial with $p(0)=0$. This last property implies that $H(0) = 0$, so that the local limit (i.e. $\alpha q^2 \rightarrow0$) is recovered correctly. The asymptotically polynomial high energy behavior of \eqref{ff} makes the corresponding nonlocal theory of gravity power-counting superrenormalizable and makes it a candidate for a putative UV completion of gravity.\\
Another widespread entire form factor choice is 
\begin{equation}
    H(\alpha q^2) = (\alpha q^2)^n.
\end{equation}
The Wataghin form factor case $n=1$ \cite{Wataghin:1934ann} is ubiquitous in effective actions of level truncated string field theories \cite{Kostelecky:1989nt,Ohmori:2001am} and, crucially, analytically tractable. The latter is the reason why it will be our benchmark case throughout the paper. 
At the level of effective theories, also non-analytic form factors of the schematic form
\begin{equation}
    \omega(\Box)=\dfrac{1}{\Box^k},\quad k=1,2
\end{equation}
proven to be useful in cosmological scenarios \cite{Barvinsky:2011hd,Deser:2007jk,Maggiore:2014sia}, due to their natural IR-regularization.\\
For our purposes we add to \cref{starting_action} a massless scalar field minimally coupled to gravity, whose action is
\begin{equation}\label{nonlocal_massless_scalar_action}
    S_{s} = - \frac{1}{2}\int d^Dx\sqrt{|g|}\Big(g^{\mu\nu}\partial_{\mu}\phi \partial_{\nu} \phi\Big).
\end{equation}
It is important to notice that one may have also included a scalar form factor in the kinetic term, but it is possible to realize that it can be removed from the leading $\phi\phi h$ vertex by a field redefinition \cite{Anselmi:2006yh,Dona:2015tra,Modesto:2021soh}, so that it will not play any role, at least at leading order and for form factors with a well defined  Jacobian. Indeed, the vertex can be computed starting from the action with a modified minimal coupling
\begin{equation}
    S_{s} =  \frac{1}{2}\int d^Dx\sqrt{|g|}\Big(\phi \Box f(-\alpha_s\Box)\phi\Big),
\end{equation}
for which it can be shown that on shell  it reduces to the usual local vertex.
This feature implies that when one considers this specific gravitational process the nonlocal effects are completely encoded in the nonlocality  of the graviton. 
\\
The  propagator for the scalar field  turns out to be  the usual one
\begin{equation}
    G(p^2) = -i\frac{1}{p^2 -i0}.
\end{equation}
It is easy to check that the $\phi\phi h$ vertex, relevant for the tree level Born exchange amplitude, is not affected by the nonlocality in the gravitational sector, and coincides with the usual vertex, with all momenta outgoing
\begin{equation}
    \tau_{\text{loc}}^{\mu\nu}(p_1,p_2) = -i\kappa_D\big(p_1^\mu p_2^\nu + p_1^\nu p_2^\mu - \eta^{\mu\nu}p_1p_2\big). 
\end{equation}
With the usual definition of the S matrix (see Appendix \ref{app:A} for other conventions)
\begin{eqnarray}
    S=1+iT,
\end{eqnarray}
 and taking the massless scalars as asymptotic states with all external momenta outgoing, we shall employ the following definition of the amplitude $\mathcal{A}(s,t)$ :
\begin{equation}
   \langle p_4,\,p_3|\,iT\,| -p_2,\,-p_1\rangle = i(2\pi)^D \delta^{D}(p_1+p_2+p_3+p_4)\Amplitude(s,t),
\end{equation}
where the Mandelstam invariants and the momentum transfer $q$ are chosen in agreement with the convention 
\begin{equation}
    s=-(p_1+p_2)^2, \quad t=-(p_1+p_4)^2 \equiv-q^2.
\end{equation}
If the exponentiation  mechanism mentioned in the introduction  holds, the leading eikonal  $2\delta_0$  is extracted from the following normalized Fourier transform to impact parameter space
 \begin{equation}
    2i\delta_0(s,b) = i \tilde{\Amplitude}_0(s,b) \equiv  i \int \frac{d^{D-2}q}{(2\pi)^{D-2}} e^{ib\cdot q} \frac{\Amplitude_0(s,-q^2)}{2s}.
\end{equation}
where we denote by $\Amplitude_{n-1}(s,t) $ the $(n-1)$ loop amplitude involving $n$ graviton exchanges between the two energetic worldlines. In particular, $\Amplitude_0(s,t)$ is the tree level amplitude. 
At tree level, as usual, the Regge limit $-s/t\rightarrow \infty$ at fixed $t$ selects the $t$-channel exchange diagram as the dominant one.
\subsection{Gaussian form factor: degenerate strings?}
To get a flavour of the effects of nonlocality, we start with massless scalar scattering, whose study was partly anticipated in \cite{Giaccari:2018nzr} in connection with macrocausality issues. In this specific case the exponentiation proceeds exactly as usual, as the vertices are left unmodified and the internal graviton propagators enter entirely as spectators in the combinatorics, as we show in  Appendix \ref{app:B}.\\
The Regge limit of the Born exchange amplitude is easily computed to be

\begin{equation} \label{born}
    i\Amplitude_{0}(s,t=-q^2) = i\kappa_D^2s^2 \frac{e^{-H(\alpha q^2)}}{q^2},
\end{equation}
from which we extract the eikonal
\begin{equation}\label{delta_0}
    2i\delta_0(s,b)  = i \frac{\kappa_D^2 s}{2}\int \frac{d^{D-2}q}{(2\pi)^{D-2}} e^{ib\cdot q}  \frac{e^{-H(\alpha q^2)}}{q^2}.
\end{equation}
For concreteness, let us now consider the Wataghin form factor mentioned above
\begin{equation}
    H(\alpha q^2) = \alpha q^2,
\end{equation}
which describes a nonlocal Gaussian spreading of the graviton propagator. We can  introduce a Schwinger parameter
\begin{equation}
    \int \frac{d^{D-2}q}{(2\pi)^{D-2}} e^{ib\cdot q}  \frac{e^{-\alpha q^2}}{q^2} = \int_0^{\infty}dt \int \frac{d^{D-2}q}{(2\pi)^{D-2}} e^{ib\cdot q}  e^{-(\alpha + t) q^2},
\end{equation}
and the remaining integral can be performed in terms of the incomplete gamma function
\begin{equation}
    \gamma(z,a) = \int_0^a du \:u^{z-1}e^{-u},
\end{equation}
so that

\begin{equation}\label{graviton_nonlocal_eik}
2\delta_0 (s,b) = G_D s \frac{1}{(\pi b^2)^{\frac{D-4}{2}}}\gamma\l(\frac{D-4}{2},\frac{b^2}{4\alpha}\r).
\end{equation}
It is quite interesting to compare this result with  the one for $2 \rightarrow2$ dilaton scattering in string theory, where the nonlocality is due to the extended nature of fundamental strings and parameterized by $\alpha'$.
 The eikonal \footnote{Note that we are citing the result before the eikonal \emph{operator} resummation. A detailed analysis \cite{Amati:1987wq,Amati:1987uf} shows that, in string diagrams, multi-Reggeon exchanges have the net effect of shifting the impact parameter by the difference of two averaged operator valued transverse string oscillators, capturing diffractive (tidal) excitations of the two scattering strings.}, extracted in \cite{Amati:1987uf}, reads
\begin{equation}
    2\delta_0(s,b)=G_D\,s\,
\frac{\Gamma\!\left(1-\dfrac{\alpha'}{4}\nabla_b^2\right)}
{\Gamma\!\left(1+\dfrac{\alpha'}{4}\nabla_b^2\right)}
\frac{1}{{(\pi b^2)^{\frac{D-4}{2}}}} \gamma\l(\frac{D-4}{2},\frac{b^2}{Y_c}\r),
\end{equation}
where
\begin{equation}
    Y_c = l_s^2(s) -i\pi\alpha',
\end{equation}
and
\begin{equation}
    l_s^2(s) = 2\alpha' \log \frac{\alpha's}{4}.
\end{equation}
The absence of an imaginary part in \eqref{graviton_nonlocal_eik}, instead present in the string theory result via $Y_c$, is no surprise. It is related to the production of heavy closed string states in the $s$ channel, clearly something absent in this model. Another important difference is the lack of a logarithmic enhancement of the nonlocality scale, tied to the fact that elementary strings stretch at large energies. The whole leading Regge trajectory is necessary to reproduce this feature, whereas the nonlocal field theory amplitude fails to reggeize. The simple nonlocal eikonal lacks also the ratio of Gamma functions, which can be neglected only in the $b^2 \gg l^2_s(s)$ limit. This is again related to the infinite number of poles in the string amplitude. In this regime, viewed as a function of their respective nonlocality scales, the two results share a clear similarity.
In fact, we can be more precise and identify a specific kinematic limit in which we go from the stringy result to the nonlocal field theory one. For this, it is more convenient to write the Regge limit of the string amplitude in momentum space
\begin{equation}
    \mathcal{A}_0(s,t) \sim \frac{32\pi G_D}{\alpha'} \l(-\frac{4}{\alpha't}\r)\frac{\Gamma\l(1-\dfrac{\alpha't}{4}\r)}{\Gamma\l(1+\dfrac{\alpha't}{4}\r)} \l(\frac{\alpha's}{4}\r)^2 e^{\frac{\alpha't}{2}\log\alpha's -i\pi\frac{\alpha't}{4}}.
\end{equation}
If we send $\alpha't$ to zero faster than how $\log (\alpha's)$ grows, we recover the usual, local, field theory result, in which all string-size effects vanish and GR expectations are met. If we focus instead on a correlated double scaling
\begin{align*}
    \alpha't \rightarrow0, \:\:\log\alpha's \rightarrow\infty;\\
     \:\: \\\frac{\alpha't}{2} \log(\alpha's) \equiv \alpha t  \:\: \text{fixed}
\end{align*}
then both the Gamma structure and the imaginary part degenerate, leaving behind only the effective amplitude \eqref{born} with Gaussian $e^{-H(\alpha q^2)}$.\\ 
A possible heuristic picture is the one in which each scatterer is described (in the center of mass frame) as an extremely energetic object, which would tend to spread quite a lot,  but with a tension chosen so comparatively large as to contract it at will. All this will give a fine-tuned rigid nonlocal interaction of effective size set by this interplay, $\sqrt{\alpha}$, suppressing all inelasticities. By rights, such objects hardly could still be called strings, but the limit serves as a useful illustration of how a stripped-down notion of nonlocality could play a role even in a field theoretic context. For instance, this notion could  provide a simple proxy model which could be used to study some stringy questions  in a simplified  setting, on the proviso to reinstate what is missing when a more  accurate description is needed.
\subsection{The fate of 't Hooft poles}
After obtaining the eikonal \eqref{delta_0} from the amplitude \eqref{born}, it is interesting to carry out the inverse transform after resummation, to recover the normalized S-matrix element
\begin{equation}\label{S_matrix_def}
    \mathcal{S}(s,-Q^2)=2s\int d^{D-2}b\,e^{-i\frac{b}{\hbar} \cdot Q+2i\delta_0(s,b)},
\end{equation}
with $Q$ now representing the full exchanged (macroscopic) momentum, which gets fractionated between multiple soft exchanges. For simplicity we will focus on $D=4$.\\
We recall that, in the $D=4$ Einstein-Hilbert minimally coupled to matter, the S-matrix element can be computed via dimensional regularization.
Focusing on the finite $b$-dependent leading eikonal
\begin{equation}
    2\delta_0(s,b)=-\dfrac{Gs}{\hbar}\log(\mu b)^2,
\end{equation}
where $\mu$ is a mass scale arising for dimensional reasons, the S-matrix yields 
\begin{equation}\label{local_S}
    \mathcal{S}(s,-Q^2)\simeq i \dfrac{8\pi Gs^2}{Q^2}\l(\dfrac{4\mu^2\hbar^2}{Q^2}\r)^{-i\alpha_G}\dfrac{\Gamma(1-i\alpha_G)}{\Gamma(1+i\alpha_G)},\quad \alpha_G=\dfrac{Gs}{\hbar}.
\end{equation}
Here we notice the curious so-called t'Hooft poles arising at non-positive integer arguments of the Gamma function. These singularities come from the short-distance behavior of the Coulombic interaction, relevant when scanning the small-$b$ region of the integrand in \eqref{S_matrix_def}. Replacing the t'Hooft poles values in \eqref{S_matrix_def} yields 
\begin{equation}
    \mathcal{S}(s,-Q^2)\Bigg|_{s_{n}=-i\frac{\hbar(n+1)}{G}}
    =4\pi s_n\int_0^\infty db \,b\,J_0\l(\dfrac{bQ}{\hbar}\r)(\mu b)^{-2(n+1)},\\
\end{equation}
with $J_0$ a Bessel function of the first kind.\\
The latter integral is divergent due to the small $b$ singularity of the integrand:
\begin{equation}
    b\,J_0\l(\dfrac{bQ}{\hbar}\r)(\mu b)^{-2(n+1)}\stackrel{b\into0}{\longrightarrow}b^{-(2n+1)}.
\end{equation}
These divergences are regularized in string theory \cite{Amati:1987uf} (see also \cite{Amati:1992zb}), and we expect that a similar mechanism will cancel them in any short-distance regulated theory, as in the present case.\\
As a clear example of how the regularization happens, we focus again on the Gaussian form factor in $D=4$. Being interested in the small-$b$ behaviour of the integrand, we approximate, for $b^2\ll 4\alpha$, 
\begin{equation}
 \gamma\l(z,\dfrac{b^2}{4\alpha}\r)\sim \frac{(b^2/4\alpha)^z}{z}.
\end{equation}
Recalling that 
\begin{equation}
    2\delta_0 (s,b) = \dfrac{G_D s}{\hbar} \frac{1}{(\pi b^2)^{\frac{D-4}{2}}}\gamma\l(\frac{D-4}{2},\frac{b^2}{4\alpha}\r),
\end{equation}
the eikonal is thus well described by
\begin{equation}
    2\delta_0 (s,b) \sim \dfrac{2G_D s}{(D-4)\hbar} (4\pi\alpha)^\frac{4-D}{2}\,.
\end{equation}
The leading infrared divergence in dimensional regularization $D=4-2\epsilon$ is $b$-independent as for the local case, and the regulated contribution is
\begin{equation}
   2\delta_0^{(\text{reg})} (s,b) = -\dfrac{G s}{\hbar} \ln (4\pi\alpha\mu).
\end{equation}
The small-$b$ part of the integrand for $s=s_n$ is now
\begin{equation}
    bJ_0\l(\dfrac{bQ}{\hbar}\r)(4\pi\alpha\mu)^{-(n+1)}\stackrel{b\into0}{\longrightarrow}0,
\end{equation}
giving a finite $\mathcal{S}(s_n,-Q^2)$.

\subsection{The deflection angle and generalized Aichelburg-Sexl geometries}\label{GAS}
As anticipated, the eikonal serves as a useful tool to extract classical observables. To leading order, the three main examples --- all closely tied to one another --- are the classical impulse, the scattering angle and the time delay. Throughout the paper we shall focus only on the deflection angle, extracted as follows. 
Let $Q$ be the classical macroscopic momentum exchange of the process, $ p \equiv|\vec{p}|$ be the magnitude of the initial spatial momentum in the center of mass frame, and $\Theta$ the angle between $-\vec{p}_1$ and $\vec{p}_4$. From standard kinematics:
\begin{equation}
    Q = 2 p \sin \frac{\Theta}{2}.
\end{equation}
In the leading approximation the full exchanged momentum after the eikonal resummation is given by a saddle point condition in \eqref{S_matrix_def}\footnote{We henceforth set $\hbar =1$.}
\begin{equation}
    Q^{\mu} =  \frac{\partial}{ \partial b_{\mu}} 2 \delta_{0}.
\end{equation}
Considering the magnitude of the vector $Q^{\mu}$, and remembering that gravity is attractive, for small deflection angles we can write
\begin{equation}
    \Theta \simeq -\frac{1}{p}{\partial_b} (2 \delta_0).
\end{equation}
Now, for massless particles $p\sim \sqrt{s}/2$. Applying it to \eqref{graviton_nonlocal_eik} we get
\begin{equation}
    \Theta^{(1)} = \frac{4G_D\sqrt{s}}{\pi^{\frac{D-4}{2}}b^{D-3}}\gamma\l(\frac{D-2}{2},\frac{b^2}{4\alpha}\r)\,.  
\end{equation}
As a special case, when $D = 4$,
\begin{equation}
    \Theta^{(1)} \simeq \frac{ 4 G\sqrt{s}}{b}\l(1-   e^{- \frac{b^{2}}{4 \alpha_{g}}} \r).
\end{equation}
These last results make evident the weakening effects of nonlocality on gravitational scattering.\\
 All of this connects nicely with an alternative derivation of the leading eikonal for  massless particles, carried out by 't Hooft \cite{tHooft:1987vrq}. The idea is to consider the scattering process of two particles, one of them soft and the other extremely energetic.  One then 
 studies the geodesic motion of the relatively soft particle in the Aichelburg-Sexl \cite{Aichelburg:1970dh} shock-wave produced by the extremely energetic scatterer. The relevant geometry in convenient coordinates reads
\begin{equation}
    ds^2 = -dudv + f(x_\perp)\delta(u)du^2 + dx_\perp^2, \quad \quad u=t-z,\,v=t+z\,.
\end{equation}
The metric depends on the transverse energy profile of the source entering in Einstein's equations. Indeed, for a pointlike distribution in transverse space, moving with energy $E^{(1)}$ and  localized at $u=0$ , the function $f(x_\perp)$ must satisfy
\begin{equation}
\label{green}
    \Box_\perp f(x_\perp) = -16\pi G_D E^{(1)}\delta^{(D-2)}(x_\perp).
\end{equation}
A second particle moving in this geometry with energy $E^{(2)}$, at $v=0$ and with transverse distance $b$ for $t<0$, will suffer a shift $\Delta v = f(b) $ at $t=0$. This will translate into a phase shift for the wavefunction of the second particle  equal to
\begin{equation}
    2\delta_0 (s,b) \equiv E^{(2)} \Delta v = E^{(2)}f(b).
\end{equation}
The leading eikonal in Einstein gravity and delta-like sources is recovered by solving \eqref{green} in terms of the Green function in the $(D-2)$ dimensional transverse space:
\begin{align}
    2\delta_0(s,b) &= 4E^{(1)}E^{(2)}G_D \frac{1}{(\pi b^2)^{(D-4)/2}}\Gamma\l(\frac{D-4}{2}\r) =\\
    &=G_Ds\frac{1}{(\pi b^2)^{(D-4)/2}}\Gamma\l(\frac{D-4}{2}\r),
\end{align}
where we have used the fact that $4E^{(1)}E^{(2)} = s$. This is indeed the local limit of \eqref{delta_0}.
As we turn on nonlocality ---  hence form factors --- in the action, we expect the relevant effective geometry experienced by the probe to change. The eikonal derived from the amplitude \eqref{born} is connected, as before, to $f(x_\perp)$ by
\begin{equation}\label{EF}
    2\delta_0(s, x_\perp) = E^{(2)}f(x_\perp),
\end{equation}
but now we generically expect  nonlocality to generate spreading along the perpendicular direction so that  $f(x_\perp)$ will not have a simple $\delta$-like support as in \eqref{green}. We are therefore led to the Ansatz
\begin{equation}\label{boxf}
    \Box_\perp f(x_\perp) = -16\pi G_D \rho^{(1)}(x_\perp),
\end{equation}
in order to ensure that  $f(x_\perp)$ describes the effects of a generalized (effective) energy density profile, which we aim to link with the form factors appearing in the theory. Indeed, combining \eqref{EF}, \eqref{boxf} and \eqref{delta_0}, we can read the effective energy density profile
\begin{equation}
    \rho^{(i)} (x_\perp) =E^{(i)} \int \frac{d^{D-2}q}{(2\pi)^{D-2}} e^{i q\cdot x_\perp} e^{-H(\alpha q^2)}.
\end{equation}
The local limit correctly yields
\begin{equation}
    \rho^{(i)}(x_\perp) \rightarrow_{\alpha =0} E^{(i)} \delta^{(D-2)}(x_\perp),
\end{equation}
while the Gaussian form factor, unsurprisingly, gives a Gaussian profile
\begin{equation}
    \rho^{(i)} (x_\perp) = E^{(i)} \frac{1}{\l(2\sqrt{\pi}\ell\r)^{D-2}}e^{-\dfrac{x_{\perp}^2}{4\ell^2}}.
\end{equation}
Other analytic form factors yield qualitatively similar results whose closed form is not particularly illuminating.

\section{Massive scalar scattering: probe limit}\label{Massive scalar scattering}
Let us now generalize our discussion on the eikonal to the case of the gravitational scattering of two minimally coupled massive scalars fields of mass $m_1,m_2$. For the first scalar field the action now reads
\begin{equation}\label{nonlocal_massive_scalar_action}
    S_{m} = - \frac{1}{2}\int d^Dx\sqrt{|g|}\Big(g^{\mu\nu}\partial_{\mu}\phi \partial_{\nu} \phi-m_1^2\phi^2\Big),
\end{equation}
and similarly for the second.
In the Regge limit $q^2\ll s, m_1^2,m_2^2$, the tree level  amplitude dressed with the form factor becomes
\begin{equation}\label{massiveBorn}
     \Amplitude_{0} (s,t = - q^{2}) = 2\kappa^2_D\frac{e^{-H(\alpha q^2)}}{q^{2}}\l[ \frac{(s - m_{1}^{2} - m_{2}^{2})^2}{2} - \frac{2 m_{1}^{2} m_{2}^{2}}{D-2} - q^{2} \frac{s - m_{1}^{2} - m_{2}^{2}}{2} \r] \, .
\end{equation}
Focusing on the terms with the strongest $q^2$ singularity \footnote{In GR and other local theories, the last term in \eqref{massiveBorn} is usually discarded on account of the fact that it would yield contributions localized in impact parameter space. Here, the same argument does not hold due to the presence of the form factor, but we discard it anyway as it is subleading compared to the combination of the other two. Reformulating the equation in terms of $(\sigma,\,m_1,\,m_2)$, the third term is  suppressed by $q^2/ m_1m_2\ll1$, given the typical hierarchy of scales $1/m \ll (G_Dm)^{1/{D-3}}\ll b \sim 1/q$ defining the semiclassical PM limit.}, the Born exchange amplitude simply gets dressed by a kinematic factor
\begin{equation}
    \mathcal{A}_0(\sigma,-q^2) = 2\kappa_D^2\mathcal{K}(s,m_i)\frac{e^{-H(\alpha q^2)}}{q^2},
\end{equation}
where 
\begin{equation}
    \mathcal{K}(s, m_i) = 2m_1^2m_2^2 \l(\sigma^2- \frac{1}{D-2}\r),
\end{equation}
and 
\begin{equation}
    \sigma \equiv \frac{1}{2} \l(\frac{s-m_1^2-m_2^2}{m_1 m_2}\r),
\end{equation}
is just the relative Lorentz $\gamma$ factor.
The eikonal is extracted as
\begin{equation}\label{MassiveEik}
    2\delta_0(\sigma,b) = \int \frac{d^{D-2}q}{(2\pi)^{(D-2)}} \frac{1}{4Ep} \mathcal{A}_0(\sigma,-q^2)e^{ib \cdot q},
\end{equation}
where, we recall, $E$ is the total energy in the center of mass frame and $p$ the norm of the spatial momentum of one of the two incoming scatterers in the center of mass frame.\\
The analysis with generic masses closely parallels the massless case, so we specialize quickly to a particular hierarchy of scales which will allow us to explore black hole physics in nonlocal theories. 
\subsection{Probe limit and eikonal compatibility}

A very instructive regime of the process is described by the \emph{probe limit}. The latter  describes  the situation in which the mass $m_2 \equiv M$ of one of the two bodies is far greater than the total energy of the other one, in the frame in which the heavy particle is at rest.
In this limit the various kinematic quantities in \eqref{MassiveEik} degenerate as follows:
\begin{align}
    E \rightarrow &\sqrt{M^2+2ME_1},\\
    p \rightarrow& \sqrt{E_1^2-m_1^2},\\
    \sigma \rightarrow& \frac{E_1}{m_1}.
\end{align}
This is precisely the regime in which one can make contact with the physical scenario of one probe particle of mass $m_1$ and energy $E_1$ moving in the background geometry produced by the heavy particle. The leading eikonal becomes
\begin{equation}
    2\delta_0  = \frac{\kappa_D^2M}{\sqrt{E_1^2-m_1^2}}\l[\frac{(D-2)E_1^2-m_1^2}{D-2}\r] \int \frac{d^{D-2}q}{(2\pi)^{D-2}}e^{iq \cdot b} \frac{e^{-H(\alpha q^2)}}{q^2}.
\end{equation}
The leading PM deflection angle, extracted again by saddle point methods, reads
\begin{equation}
    \Theta^{(1)} = -\frac{1}{p} \partial_b (2\delta_0).
\end{equation}
One can perform the angular integration in the Fourier transform and the derivative, with the result
\begin{equation}
    \Theta^{(1)} = \frac{4G_DM}{b^{D-3}}\l[\frac{(D-2)E_1^2-m_1^2}{(D-2)(E_1^2-m_1^2)}\r]\frac{1}{(2\pi)^{(D-4)/2}}  \int_0^\infty d\gamma\: \gamma^{\frac{D-4}{2}}J_{\frac{D-2}{2}}(\gamma)e^{-H(\alpha\gamma^2/b^2)},
\end{equation}
where $J_{(D-2)/2}(\gamma)$ is a Bessel function.
Further using the definition of the Schwarzschild radius given in \eqref{Rs}, we can reduce the result to
\begin{align} \label{eiktheta}
    \nonumber
    \Theta^{(1)} &= \chi(D) \l(\frac{R_s}{b}\r)^{D-3}\l[\frac{E_1^2-\frac{m_1^2}{(D-2)}}{E_1^2-m_1^2}\r]\int_0^\infty d\gamma \:\gamma^{\frac{D-4}{2}}J_{\frac{D-2}{2}}(\gamma)e^{-H(\alpha\gamma^2/b^2)}\\
    &\equiv \chi(D) \l(\frac{R_s}{b}\r)^{D-3}\l[\frac{E_1^2-\frac{m_1^2}{(D-2)}}{E_1^2-m_1^2}\r]\mathcal{I}[H](b^2/\alpha),
\end{align}
where for notational simplicity we defined
\begin{equation}
    \chi(D) \equiv \frac{(D-2)\sqrt{\pi}}{2^{\frac{D-2}{2}}\Gamma\l(\frac{D-1}{2}\r)},
\end{equation}
and  $\mathcal{I}[H](b^2/\alpha)$  is defined by the last line of the previous equation. As a simple sanity check, we set $H=0$ and $m_1=0$ in \eqref{eiktheta}. We obtain the result
\begin{equation}
    \Theta^{(1)} = \sqrt{\pi} \frac{\Gamma\l(\frac{D}{2}\r)}{\Gamma\l(\frac{D-1}{2}\r)}\l(\frac{R_s}{b}\r)^{D-3},
\end{equation}
which  agrees nicely with the known 1PM result for the motion of a massless probe in a $D$ dimensional Schwarzschild black hole metric \cite{KoemansCollado:2019ggb}, as it should. So does the case for $m_1 ^2 \neq 0$.\\
Let us now specialize, as done in the massless case, to $H(t) = t$, the simple Gaussian form factor. Direct integration yields
\begin{equation}\label{thetaWat}
    \Theta^{(1)} = \l[\frac{(D-2)E_1^2 -m_1^2}{E_1^2-m_1^2}\r]\frac{\sqrt{\pi}}{2\Gamma\l(\frac{D-1}{2}\r)}\gamma \l(\frac{D-2}{2},\frac{b^2}{4\alpha}\r)\l(\frac{R_s}{b}\r)^{D-3}.
\end{equation}
In the large nonlocality limit $\alpha \equiv\ell^2\gg b^2$ we get 
\begin{equation}
    \Theta^{(1)} \approx \l[\frac{E_1^2 - \frac{m_1^2}{D-2}}{E_1^2-m_1^2}\r] \frac{\sqrt{\pi}}{\Gamma\l(\frac{D-1}{2}\r)}\frac{R_s^{D-3}b}{(2\ell)^{D-2}}.
\end{equation}
As expected the deflection angle vanishes linearly  with the impact parameter $b$, signaling the weakening effect that nonlocality has on gravitational interactions.\\
\begin{figure}[t]
  \centering
  \includegraphics[
  width=0.73\textwidth]{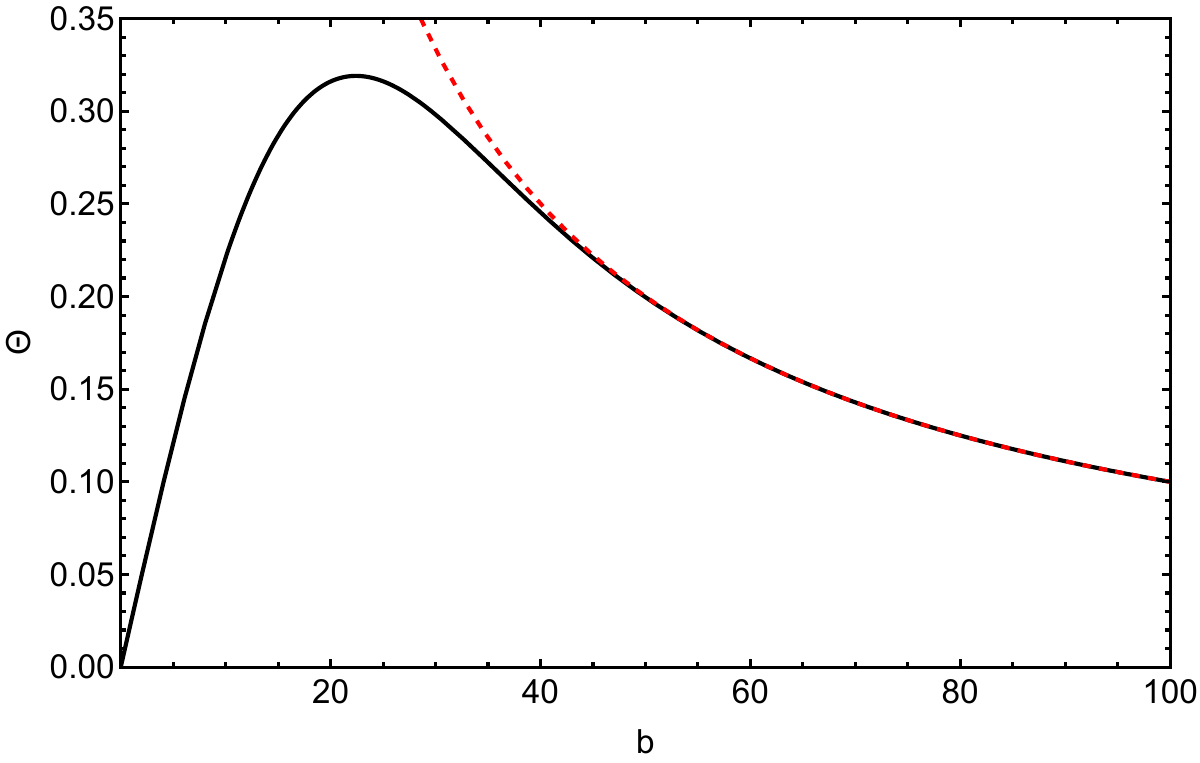}
  \caption{\label{DeflectionA} The leading PM  deflection angle $\Theta^{(1)}$ in $D=4$ for our nonlocal gravity theory with a Gaussian form factor  and for a massless particle in the probe limit (solid black line), as a function of the impact parameter $b$. In red, the famous Einstein result $4GM/b$. Units are chosen so that $\ell=10$, and we fix $2GM=5.$}
\end{figure}Setting $D=4$ and $m_1=0$ we  get  from \eqref{thetaWat}
\begin{equation}
    \Theta^{(1)} = \frac{2R_s}{b}\l(1-e^{-\frac{b^2}{4\alpha}}\r),
\end{equation}
shown in  Figure \ref{DeflectionA} and in agreement with known results \cite{Buoninfante:2020qud} for the motion of massless probes in a linearized \emph{smeared} $D=4$ Schwarzschild metric, whose form in isotropic coordinates is given by
\begin{equation}\label{smeared_schw}
    ds^2 = -(1+2\Phi)dt^2 + (1-2\Phi)\l(dr^2 +r^2 d\Omega_2\r).
\end{equation}
Here $\Phi$ is the nonlocality-corrected Newtonian potential, which can also be extracted via the usual Fourier transform of the graviton propagator and takes the form
\begin{equation}
    \Phi(r) = - \frac{GM}{r} \text{Erf}\l(\frac{r}{2\ell}\r), \:\:\:\: \ell \equiv\sqrt{\alpha}.
\end{equation}
Notice that, as expected,  the form factor regularizes the $r=0$  divergence  of the GR  Newtonian potential.

\section{Reconstruction of a nonlinear form of the metric}
\label{sect:nonlinear}
In the previous section we have used the deflection angle computed from the eikonal  approximation of the nonlocal gravitational scattering of scalars to reconstruct the form of the background gravitational metric at linearized (1PM) level. This is possible because in the probe limit the gravitational scattering of the two scalars  can be seen as the geodesic motion of the light particle (the probe) in the fixed  gravitational field generated by the heavy particle.   The effects of nonlocality in the propagator  for  the gravitons generate in this way an effective Newtonian potential, which can be used to write down the spacetime metric  at linearized level. In principle,  the result obtained in this way cannot be trusted beyond the range of validity of the linearized 1PM approximation.  
On the other hand  going beyond the linearized solution has proven to be a formidable  problem in these theories \cite{Buoninfante:2022ild}, given the complicated character of the full equations of motion. Some progress in this direction was made when Nicolini and Spallucci \cite{Nicolini:2005vd}, motivated by noncommutative geometry, derived a simple and elegant singularity-free metric for a spherically symmetric black hole. Later, in \cite{Modesto:2010uh}, it was realized that such a metric could be an approximate solution of nonlocal gravity \cite{Modesto:2011kw} with schematic equations of motion
\begin{equation}
e^{H(\Box)} G_{\mu\nu} + \mathcal{O}({\rm Riem}^2) = 8 \pi G_D \, T_{\mu\nu} \, .
\label{effective}
\end{equation}
Moving the nonlocality on the right hand side, neglecting the terms $O({\rm Riem}^2) $ and introducing an effective stress-energy tensor describing an anisotropic fluid we get:
\begin{equation}\label{effective_EOM}
G_{\mu\nu} \simeq 8 \pi G_D \, e^{ - H(\Box)} T_{\mu\nu}\equiv  8 \pi G_D
T_{\mu\nu}^{\rm eff}
\, .
\end{equation}
In order to solve the above equations of motion we need to specify the equation of state for the anisotropic fluid or to consider the linearized equations of motion \cite{Buoninfante:2022ild}. 
If we follow the former path, the simplest choice consists in assuming the Schwarzschild form of the metric, namely $g_{tt} = - g_{rr}^{-1}$, which in turn implies:
\begin{equation}\label{pmr}
    p_r = - \rho. 
\end{equation}
According to the above choice, we get the original Spallucci-Nicolini metric \cite{Modesto:2010uh}, whose explicit form generalized to $D$ dimensions reads
\begin{align} \label{NM} \nonumber
    ds^2= &-\l[1-\l(\frac{R_s}{r}\r)^{(D-3)}\frac{\gamma\l(\frac{D-1}{2},\frac{x^2}{4}\r)}{\Gamma \l(\frac{D-1}{2} \r)}\r]dt^2 + \\ &\l[1-\l(\dfrac{R_s}{r}\r)^{(D-3)}\dfrac{\gamma\l(\frac{D-1}{2},\frac{x^2}{4}\r)}{\Gamma \l(\frac{D-1}{2} \r)}\r]^{-1}dr^2 + r^2d\Omega_{D-2},
\end{align}
 For reasons that will be clear shortly, the condition \eqref{pmr} must actually be relaxed and the geometry \eqref{NM} will have to be refined.\\
A way around the bottleneck in the solution of the equations of motion is to conjecture a form of the full nonlinear spacetime  metric using some general requirements and then constrain  it by imposing that the 1PM deflection angle extracted from geodesic motion in this metric be equal to that obtained from the eikonal phase, a consistency condition that we dub \emph{eikonal compatibility}. This procedure  should also be  equivalent to starting with the linearized metric in isotropic coordinates, whose only needed input is the Newtonian potential, then switching to standard Schwarzschild coordinates, and postulating a simple nonlinearization.\\
A quite generic expectation for the full nonlinear metric in this theory is that it lacks, thanks to the smearing property of nonlocality, the $r=0$
singularity of the Schwarzschild solution. Now, the field equations of our nonlocal gravity theory can always be rewritten in terms of Einstein equation sourced with an anisotropic fluid \cite{Cadoni:2022chn}, which gives an effective description of nonlocality effects in terms of energy density $\rho$,    
radial pressure $p_r$ and  tangential pressure $p_{\perp}$.  If one additionally assumes  the equation of state $p_r=-\rho$ for the anisotropic fluid the spacetime metric  must take the general form \footnote{We are considering the $D=4$ case.}

\begin{equation}\label{smeared_schw1}
    ds^2 = -F(r, R_s,\ell) dt^2 + F^{-1}(r, R_s,\ell)dr^2 +r^2 d\Omega_2.
\end{equation}
This form of the metric must describe a deformation of the Schwarzschild black hole solution endowed with an additional hair $\ell$ and must be free of the curvature singularity    at $r=0$.  In \cite{Cadoni:2022chn} it has been shown that if one excludes the presence of conical singularities at the origin,  the only way to remove  the curvature singularity  at $r=0$ is to  assume that near $r=0$ we have 
\begin{equation}\label{dscore}
     F= 1- \frac{\Lambda }{3}r^2 + \mathcal{O}(r^3).
\end{equation}
Depending on  the value of cosmological constant $\Lambda$, we have nonsingular black holes with dS ($\Lambda>0$), AdS ($\Lambda<0$)
or Minkowski ($\Lambda=0$) core. The cosmological constant $\Lambda$ is determined  by $R_s$ and by the  nonlocality scale $\ell$ by a precise scaling law \cite{Cadoni:2022chn}. 
Imposing quite general  assumptions about the properties of  the metric (presence of a (A)dS core,  recovery of the Schwarzschild solution in the $r\to\infty$ limit) one can show that the most general form of  the function $F$ is given by 

\begin{equation}\label{dscore1}
     F= 1- \frac{R_s}{r} h\l(\frac{r}{\ell}\r),
\end{equation}
where $h(r/\ell)$ is a function behaving as $\sim r^3$ near $r=0$ while going to $1$ for $r\to\infty$.
The previous consideration would motivate a form of the full nonlinear metric given exactly by \eqref{smeared_schw1} with $F$ of the form \cref{{dscore1}}. However, as we shall see, the eikonal compatibility condition mentioned above, for generic masses of the probe, seems to require a spacetime metric parametrized with two functions.\\
Therefore, as a line of attack, we can postulate an Ansatz for the spacetime metric which is a two-functions deformation of \eqref{smeared_schw1}
\begin{equation} \label{Ansatz}
    ds^2=-\l[1-\l(\frac{R_s}{r}\r)^{(D-3)}h(r/\ell)\r]dt^2 + \l[1-\l(\dfrac{R_s}{r}\r)^{(D-3)}g(r/\ell)\r]^{-1}dr^2 + r^2d\Omega_{D-2}\,,
\end{equation}
with $h(r/\ell)$ and $g(r/\ell)$ functions to be determined.\\
Concretely, assuming the form \eqref{Ansatz} for the metric and  that the whole $r/\ell$ dependence can be kept without spoiling the PM expansion, computing the leading deflection angle on the geometric side one finds (see Appendix \ref{app:C} for a derivation)
\begin{align} \label{geotheta}\nonumber
    \Theta^{(1)} = &\l(\frac{R_s}{b}\r)^{D-3}\frac{1}{E_1^2-m_1^2} \int_0^1 \frac{du\:u^{D-3}}{(1-u^2)^{3/2}}\bigg\{E_1^2\l[g\l(\frac{x}{u}\r) -h\l(\frac{x}{u}\r) \r] \\
    &-g\l(\frac{x}{u}\r) \l[ (E_1^2-m_1^2)u^2 +m_1^2\r] \bigg\}, \:\:\: \text{with} \:\:\: x\equiv b/\ell.
\end{align}
Equating \eqref{geotheta} and \eqref{eiktheta} for all values of $x$, for all $m_1^2$ and, independently, for all $E_1^2$ produces two integral equations for $g(x)$ and $h(x)$
\begin{align}
    \int_0^1 \frac{du}{(1-u^2)^{3/2}} \l[u^{D-1}g(x/u) - u^{D-3}g(x/u) \r] = -\frac{\chi(D)}{D-2}\mathcal{I}[H](x^2),\\
    \int_0^1 \frac{du}{(1-u^2)^{3/2}} \l\{u^{D-1}g(x/u) - u^{D-3}\l[g(x/u)-h(x/u)\r] \r\} =-\chi(D)\mathcal{I}[H](x^2).
\end{align}
Substituting the first into the second we decouple the system
\begin{align} \label{system}
    \int_0^1 du\: \frac{u^{D-3}}{(1-u^2)^{1/2}}\: g(x/u) &=\frac{\chi(D)}{D-2}\mathcal{I}[H](x^2), \\
     \int_0^1 du\: \frac{u^{D-3}}{(1-u^2)^{3/2}} \:h(x/u) &= -\chi(D) \l(\frac{D-3}{D-2}\r) \mathcal{I}[H](x^2).
\end{align}
The equations are most naturally expressed in Mellin space
\begin{align}
\nonumber
    &\tilde{g} (s) \tilde{K}_{\frac{1}{2}}(s) = \frac{\chi(D)}{D-2}\widetilde{\mathcal{I}[H]}(s),\\
    &\tilde{h} (s) \tilde{K}_{\frac{3}{2}}(s) = -\chi(D)\l(\frac{D-3}{D-2}\r)\widetilde{\mathcal{I}[H]}(s),
\end{align}
where
 \begin{align}
     \nonumber
     \tilde{K}_{\frac{1}{2}}(s) \equiv& \int_0^1 du\: u^{s-1}  \frac{u^{D-2}}{(1-u^2)^{1/2}},\\
     \tilde{K}_{\frac{3}{2}}(s) \equiv& \int_0^1 du\: u^{s-1}  \frac{u^{D-2}}{(1-u^2)^{3/2}},\\
     \nonumber
     \widetilde{\mathcal{I}[H]}(s) \equiv &\int_{0}^{\infty}dx \: x^{s-1} \mathcal{I}[H](x^2).
     \end{align}
Plugging in the explicit form of $\mathcal{I}[H]$ we extract
\begin{equation}
    \widetilde{\mathcal{I}[H]}(s) = 2^{-3+D/2 +s} \frac{\Gamma\l(\frac{D+s-2}{2}\r)}{\Gamma\l(\frac{2-s}{2}\r)} \widetilde{e^{-H}}\l(-\frac{s}{2}\r).
\end{equation}
where with $\widetilde{e^{-H}}(-s/2)$ we are indicating the Mellin transform of $e^{-H}$ evaluated at $-s/2$.
The other Mellin transforms can be defined by analytic continuation. In terms of Euler's Beta function 
\begin{equation}
    \tilde K_{c}(s) = \frac{1}{2}B\l(\frac{s+D-2}{2}, 1-c\r),\ \qquad c=\frac{1}{2},\frac{3}{2}. 
\end{equation}
The algebraic equations in Mellin space are easily solved, and after a Mellin inversion a particular solution for $g(x)$ and $h(x)$ can be found. In particular, specializing to the benchmark  example $H(t)=t$ describing the Gaussian form factor, we have
\begin{equation}
    \widetilde{e^{-H}}\l(-\frac{s}{2}\r) = \Gamma\l(-\frac{s}{2}\r).
\end{equation}
 The two particular solutions are found to be
\begin{align}   \label{hg}
    g(x) = \frac{1}{2\pi i}\int_{\mathcal{C}_1} ds \:x^{-s} \tilde{g}(s) = \frac{\gamma\l(\frac{D-1}{2},\frac{x^2}{4}\r)}{\Gamma \l(\frac{D-1}{2} \r)},\\
   h(x) = \frac{1}{2\pi i}\int_{\mathcal{C}_2} ds \:x^{-s} \tilde{h}(s) = \frac{\gamma\l(\frac{D-3}{2},\frac{x^2}{4}\r)}{\Gamma \l(\frac{D-3}{2} \r)},
    \end{align}
where $\mathcal{C}_1$ and $\mathcal{C}_2$ are vertical contours in between the nearest left and right poles of the respective integrands. Clearly, the functions above are not the unique solutions. The reason, mathematically, is that the integral operators in \eqref{system} admit a nontrivial null space, and when necessary must be defined by analytic continuation. The addition of functions belonging to these kernels will identically preserve the eikonal compatibility conditions. From a physical standpoint, it is rather obvious that many candidate metrics can yield the same 1PM deflection angle. We take the above solutions as a minimal choice, but the $g_{tt}$ so extracted is in agreement with the Newtonian potential derived from the linearized solution and this provides a consistency check.\\
Solutions of the system \eqref{system} should be completely equivalent to a doublet of functions: $h(r/\ell)$ consistent with the Newtonian potential $\Phi$ of the theory \eqref{starting_action}, and 
\begin{equation}\label{ghrelation}
    g(r/\ell)=\l(1-\dfrac{r\partial_r}{D-3}\r)h(r/\ell).
\end{equation}
Incidentally, the relation \eqref{ghrelation}, deduced from \eqref{system}, states that the linearized metric producing the correct 1PM deflection angle, when written in isotropic coordinates, is exactly of the form \eqref{smeared_schw}, naturally generalized to $D$ dimensions.
\\
The resulting metric
\begin{align} \label{trialmetric_horizonsingular} \nonumber
    ds^2= &-\l[1-\l(\frac{R_s}{r}\r)^{(D-3)}\frac{\gamma\l(\frac{D-3}{2},\frac{x^2}{4}\r)}{\Gamma \l(\frac{D-3}{2} \r)}\r]dt^2 + \\ &\l[1-\l(\dfrac{R_s}{r}\r)^{(D-3)}\dfrac{\gamma\l(\frac{D-1}{2},\frac{x^2}{4}\r)}{\Gamma \l(\frac{D-1}{2} \r)}\r]^{-1}dr^2 + r^2d\Omega_{D-2},
\end{align}
has a number of desirable properties, such as a consistent nonlocally modified Newtonian potential, presence of a  de Sitter core ensuring the absence of curvature singularity at $r=0$, eikonal compatibility, the right local limit (Schwarzschild-Tangherlini as $\ell \rightarrow0$), and a $g_{rr}$ metric component which, being naturally tied to a  mass function $M(r)$ determined from a Gaussian density, already appeared as a proposal in the literature (see e.g. \cite{Modesto:2010uh}).\\
It is also important to note that the Ansatz we put forward was motivated by regularity, as it is essentially a \emph{dirty} (in the sense of \cite{Visser:1992qh} of having $-g_{tt}\neq g^{-1}_{rr}$) deformation of the regular black hole metrics studied in a number of works \cite{Cadoni:2022chn, Cadoni:2022vsn, Hayward:2005gi, Nicolini:2005vd}, expected to play a key role in consistent UV completions of gravity. The necessity of a dirty solution, as opposed to a ``clean" one with $-g_{tt}= g^{-1}_{rr}$, stems from the  difficulty mentioned above of finding a simultaneous solution of \eqref{system} for \emph{non-constant} $h(x)=g(x)$. The system may collapse in this case to an overconstrained one, perhaps hinting at a deeper physical reason for an obstruction of this type.\\ The form of the spacetime is strongly related to the equation of state satisfied by the effective anisotropic fluid sourcing the metric solution.  It is well-known that a ``clean" black hole  must be sourced by  fluid with EOS $p_r =-\rho$. This means that our dirty black hole solution must be sourced by a fluid with the following EOS
\begin{equation}\label{EOS}
    p_r \neq -\rho.
\end{equation}
In addition, nonlocality is widely believed to be a possible mechanism to fully smear singularities in gravitational theories \cite{Koshelev:2024wfk,Buoninfante:2018xiw,Edholm:2016hbt,Koshelev:2018hpt,Burzilla:2020utr,Giacchini:2016xns}. In fact the geometry in \eqref{trialmetric_horizonsingular} removes the central singularity of the Schwarzschild solution. Unfortunately, a closer inspection shows that, while being perfectly regular at $r=0$ thanks to an emergent de Sitter core, it is nonetheless singular at the Killing horizons determined by $g_{tt}=0$ (if the parameters are such that they exist).
Calculating the scalar curvature $R$ for a generic ``dirty" metric one can easily show that a zero of $g_{tt}$ is a curvature singularity unless it is also a zero of $g_{rr}^{-1}$.  This in particular implies that the scalar curvature of the spacetime diverges at the event horizon (if existing) of our solution \eqref{trialmetric_horizonsingular}.
In the next section we will see how one can solve this problem.  We will find an everywhere regular black hole solution.
\subsection{The  nonlinear completion }\label{guess}
We are now finally in a position to sharpen our Ansatz, guided by the basic and physically motivated requirements presented at the beginning of this section, namely:  eikonal compatibility and absence of curvature singularities in the \emph{whole} spacetime.
To enforce the last condition at the possible horizons, we look for a $g_{tt}$ sharing the same zeroes as $g_{rr}^{-1}$ in \eqref{trialmetric_horizonsingular}. A suitable Ansatz satisfying this condition by construction is
\begin{equation}\label{sigma_gtt}
    g_{tt} =- \Sigma(R_s/r, \ell/r) \l[ 1 - \l(\dfrac{R_{s}}{r}\r)^{D-3}g(x)\r], \qquad g_{rr}^{-1}= -\frac{g_{tt}}{\Sigma},
\end{equation}
where $\Sigma(r/\ell, R_s/r)$ is nowhere vanishing, positive definite, and admits a PM expansion
\begin{equation} \label{sigma_non_exp}
    \Sigma(x\equiv r/\ell, R_s/r) \approx 1-\l(\frac{R_s}{r}\r)^{D-3}\bar{h}(x)  + \mathcal{O} \l[\l(\frac{R_s}{r}\r)^{2(D-3)} \r].
\end{equation}
Then, to 1PM order
\begin{equation}
    -g_{tt} = { 1 - \l(\dfrac{R_{s}}{r}\r)^{D-3}\l[\bar{h}(x) +g(x)\r] +  \mathcal{O} \l[\l(\frac{R_s}{r}\r)^{2(D-3)} \r]},
\end{equation}
and by the eikonal compatibility condition spelled out in the previous section, which is here equivalent to the requirement of the correct Newtonian potential, we must have $\bar{h}(x) +g(x) = h(x)$, so that $\bar{h}(x) = h(x) - g(x)$. Using \eqref{hg} and $\gamma$ recursion relations we find
\begin{equation}
    \bar{h}(x) =  \l(\frac{x}{2}\r)^{D-3} \dfrac{e^{-\frac{x^2}{4}}}{\Gamma\l(\frac{D-1}{2}\r)}.
\end{equation}
With this function at hand, it is now natural, though of course conjectural, to promote the expansion \eqref{sigma_non_exp} to a full-fledged resummed exponential 
\begin{equation}\label{sigmaexp}
    \Sigma(x,R_s/r) = e^{-\bar{h}(x)\l(\frac{R_s}{r}\r)^{D-3}}.
\end{equation}
As a justification of this step, we note that it is rather common to parameterize dirty solutions with exponentiated redshift functions \cite{Visser:1992qh}. 
Had we started directly with the Ansatz in exponentiated form, our procedure would anyway have led to an identical expression for the exponent.
The resulting geometry, described by
\begin{align} \label{trialmetric} \nonumber
    ds^2= &-\:e^{-\l(\dfrac{R_s}{2\ell}\r)^{D-3} \dfrac{e^{-\frac{r^2}{4\ell^2}}}{\Gamma\l(\frac{D-1}{2}\r)}}\l[{1-\l(\dfrac{R_s}{r}\r)^{(D-3)}\dfrac{\gamma\l(\frac{D-1}{2},\frac{r^2}{4\ell^2}\r)}{\Gamma \l(\frac{D-1}{2} \r)}}\r]dt^2 + \\ & + \l[1-\l(\dfrac{R_s}{r}\r)^{(D-3)}\dfrac{\gamma\l(\frac{D-1}{2},\frac{r^2}{4\ell^2}\r)}{\Gamma \l(\frac{D-1}{2} \r)}\r]^{-1}dr^2 + r^2d\Omega_{D-2},
\end{align}
is indeed regular everywhere and a thorough study to check it, involving the curvature invariants $R,\, R_{\mu\nu}R^{\mu\nu}, R_{\mu\nu\rho\sigma}R^{\mu\nu\rho\sigma}$, was performed in $D=4$. As an example, at the origin we have
\begin{equation}
    R(0)=\dfrac{GM}{\sqrt{\pi}\ell^3},\quad R_{\mu\nu}R^{\mu\nu}(0)=\dfrac{G^2 M^2}{\pi\ell^6}, \quad R_{\mu\nu\rho\sigma}R^{\mu\nu\rho\sigma}(0)=\dfrac{5}{3}\dfrac{G^2 M^2}{\pi\ell^6}.
\end{equation}
Moreover, in four dimensions it shares many features with the non-commutativity-inspired dirty black hole metric  studied in \cite{Nicolini:2009gw}. 
This was perhaps to be expected, as it is known that nontrivial commutation relations between coordinates effectively induce a minimal length, which can be identified with the nonlocality scale. Pointlike distributions in these theories must be replaced with minimal width Gaussians, hence the overlap with our complementary, albeit technically very different, analysis.    
\\
Near the origin we find a de Sitter core with a rescaled time coordinate, which is what concretely replaces the singularity
\begin{align}\nonumber
    ds^2 \approx\: & -e^{-\l(\dfrac{R_s}{2\ell}\r)^{D-3} \dfrac{1}{\Gamma\l(\frac{D-1}{2}\r)}}\l[1- \frac{R_s^{D-3}}{\Gamma\l(\frac{D+1}{2}\r) (2\ell)^{D-1}}r^2\r]dt^2 +\\
    & + \l[1- \dfrac{R_s^{D-3}}{\Gamma\l(\frac{D+1}{2}\r) (2\ell)^{D-1}}r^2\r]^{-1}dr^2 + r^2 d\Omega_{D-2}, \qquad r\ll \ell\,,
\end{align}
with an effective cosmological constant given by
\begin{equation}
    \Lambda_{\text{dS}} = \frac{D-2}{\Gamma \l(\frac{D-1}{2}\r) } \frac{R_s^{D-3}}{(2\ell)^{D-1}}.
\end{equation}
In conclusion to this section, it is worth noting that \eqref{EOS} holds again. The latter is a distinctive feature of dirty black holes, whose specific EOS can be derived 
from Einstein's equations \eqref{effective_EOM} expressed in terms of an (effective) anisotropic fluid with stress-energy tensor 
\begin{equation}
     (T^{\mu}_{\,\,\nu})^{\rm{eff}}=\mathrm{diag}\l[\rho(r),\,p_r(r),\,p_{\perp}(r),\,p_{\perp}(r)\r],
\end{equation}
yielding \cite{Nicolini:2009gw}
\begin{align}
 \frac{ d M}{dr} &= 4 \pi \rho \, r^2 ,  \nonumber \\
 \frac{1}{g_{tt}} \frac{ d g_{tt}}{dr} &= \frac{2G \left( M(r) + 4 \pi  p_r  r^3\right)}{r (r-2G M(r))} ,  \\\nonumber
  \frac{d p_r}{d r} &= - \frac{1}{g_{tt}}  \frac{ d g_{tt}}{d r} (\rho + p_r) + \frac{2}{r} (p_{\bot} - p_r ).
\label{eeq}
\end{align}
The most relevant properties of the anisotropic fluid accounting for \eqref{trialmetric} are:

\begin{itemize}
    \item A Gaussian energy density responsible for the $M(r)$ in the $g_{rr}$ component, as for \eqref{trialmetric_horizonsingular};
    \item The tangential and (always negative) radial pressures behave smoothly and vanish asymptotically;
    \item The effective energy-momentum tensor is, trivially, covariantly conserved;
    \item Regularity of $T^{\mu}_{\,\,\nu}$ components emerged as a consequence of the proposed procedure, as opposed to \cite{Nicolini:2009gw} where some of the previous features were the outcome of a suitably chosen EOS relating $\rho$ and $p_r$.
\end{itemize}
\subsection{Geometric and thermodynamical properties of the solution}
\label{sect:geometric}
\begin{figure}[t]
    \centering
    \includegraphics[width=0.73\linewidth]{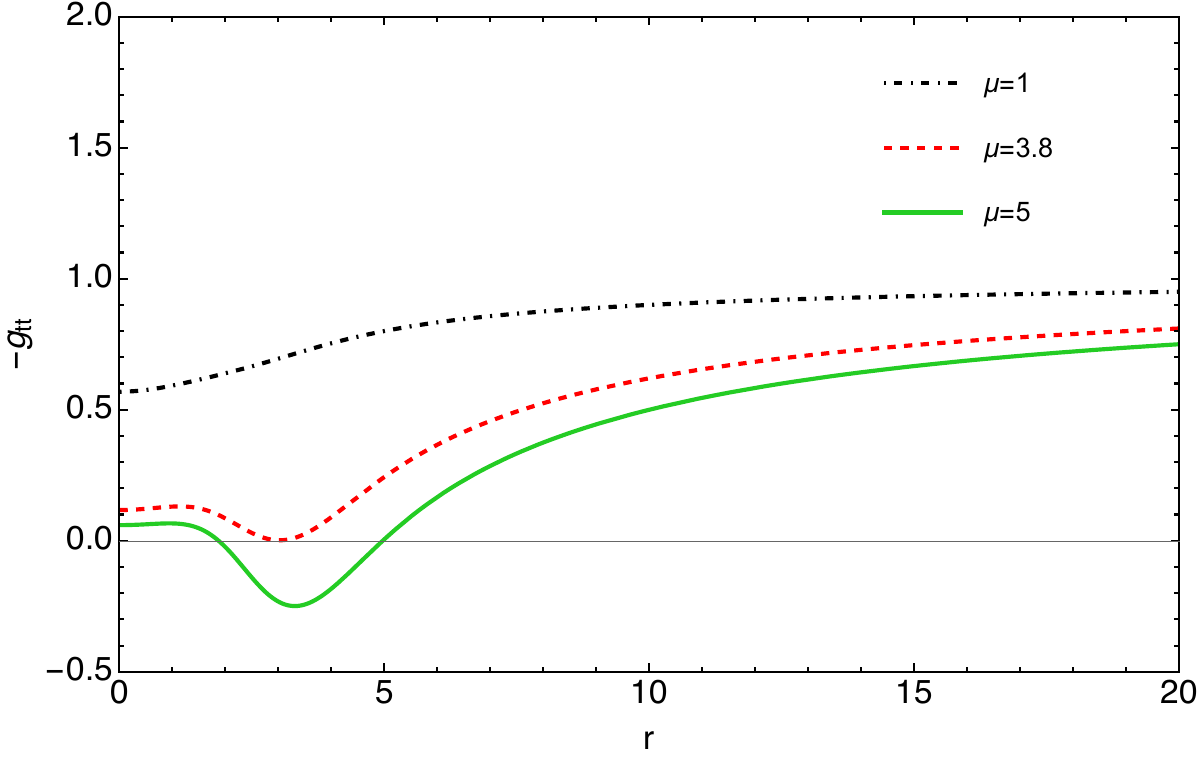}
    \caption{\label{fR_graph} Plot of $-g_{tt}$ vs $r$ for \eqref{trialmetric} in $D=4$. A different number of horizons is obtained depending on $\mu$. Specifically, we here observe zero horizons (black dashdotted line, $\mu=1$), one extremal horizon (red dashed line, $\mu\sim3.8$) and two horizons (green solid line, $\mu=5$) in units where $\ell=1$. Notice that the horizontal position of the minimum depends on $\mu$ because of the exponential prefactor \eqref{sigmaexp}. The $r=0$ value is again dictated by the exponential, whereas the large $r/\ell$ limit is $g_{tt}=1$ in order to recover Schwarzschild.}
\end{figure}

\begin{figure}[t]
    \centering
    \includegraphics[width=0.49\textwidth]{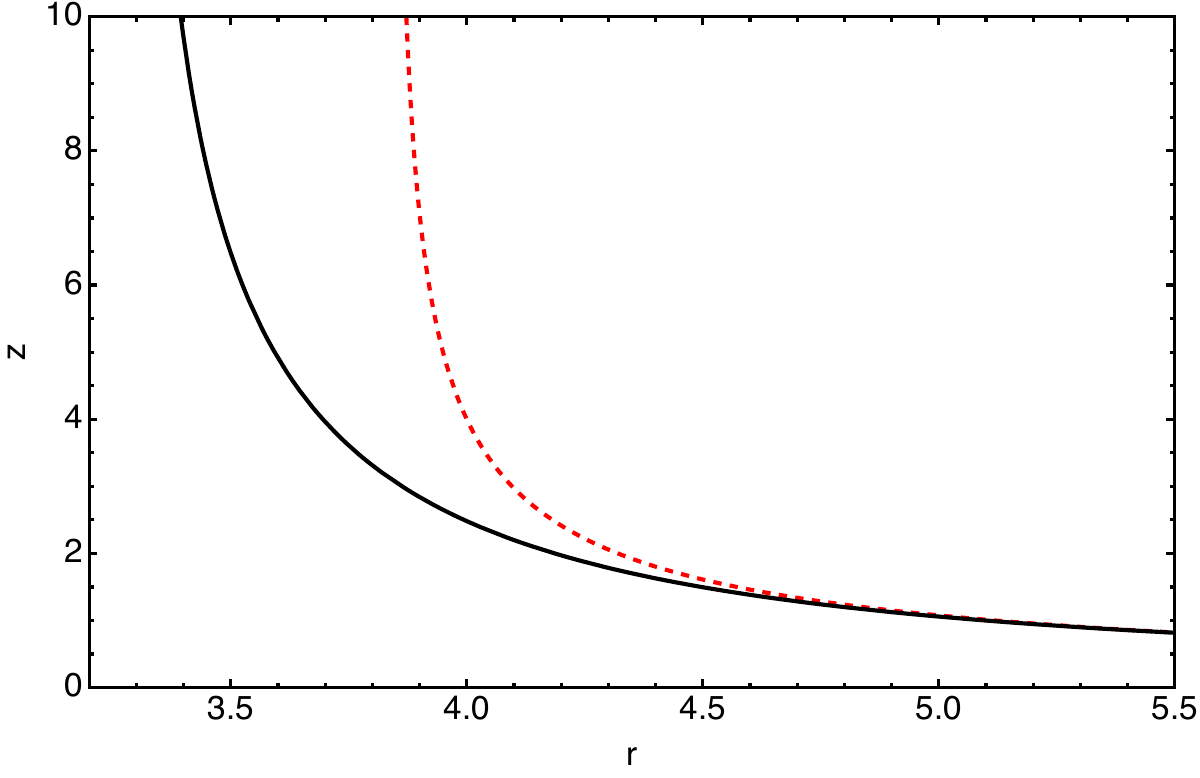}
    \hfill
    \includegraphics[width=0.479\textwidth]{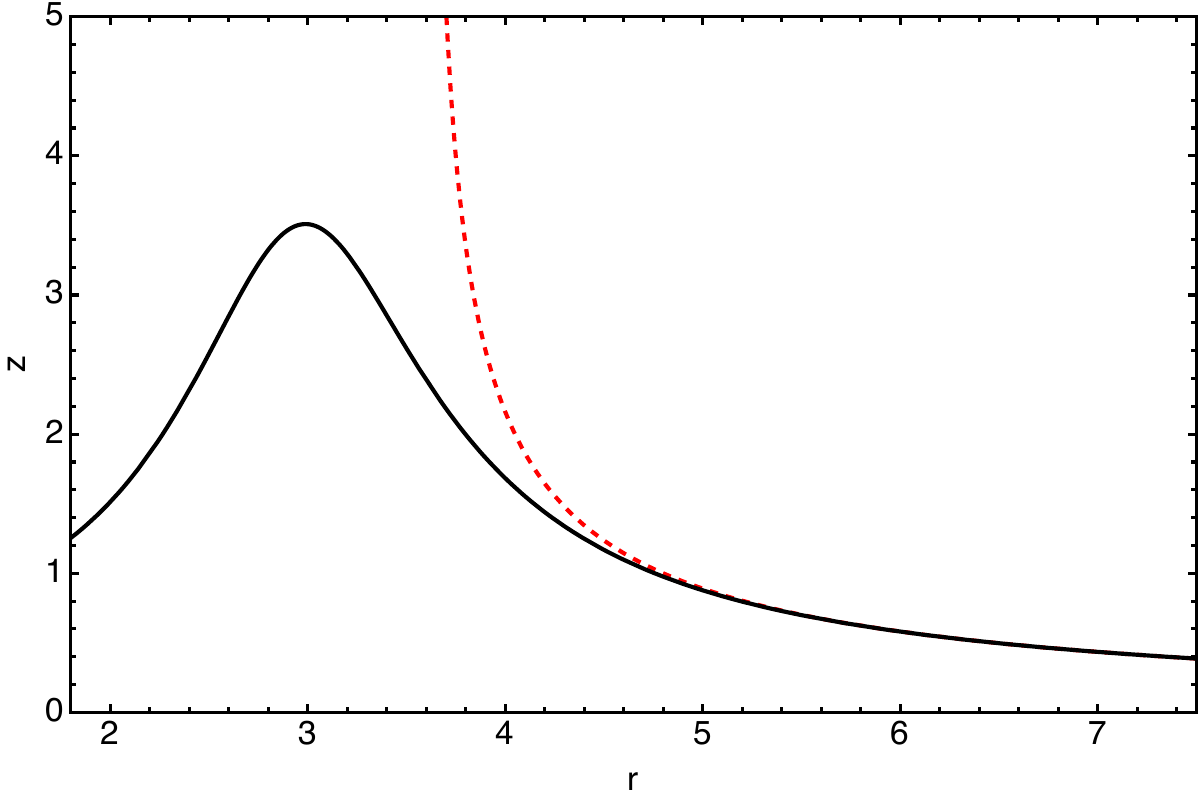}
    \caption{\label{redshift_graph}
    \textbf{Left}: Comparison of the asymptotic redshift between the Schwarzschild case (dashed red line) and the dirty nonlocal one \eqref{trialmetric} (black line) for the case with \emph{two horizons}. We set $R_S=3.84\,\ell$ in units where $\ell=1$. The extremal case is quite similar. Notice the divergence of both redshift functions when approaching the respective horizons $r_{+}\sim 3.28\,\ell$ and $r_{\rm{Schw}}=R_S$. \textbf{Right}: Comparison in the \emph{horizonless} case, where we chose $R_S=3.6\,\ell$. The redshift \eqref{redshif} is now regular everywhere and peaked around the would-be extremal horizon $r=r_0$.}
\end{figure}

In order to assess the nature of the effective compact object described by \eqref{trialmetric}, it is crucial to study the Killing horizons of the metric, defined by the solutions of the equation $g_{tt}=0$, the outer of which, for our static solution, coincides with the event horizon. We have
\begin{equation}\label{fR}
  f(r)\equiv \l[{1-\mu \l(\dfrac{\ell}{r}\r)^{(D-3)}\dfrac{\gamma\l(\frac{D-1}{2},\frac{x^2}{4}\r)}{\Gamma \l(\frac{D-1}{2} \r)}}\r]=0, \quad \l(\dfrac{R_s}{\ell}\r)^{D-3}\equiv\mu. 
\end{equation}
It is clear, in analogy to elsewhere studied $D=4$ regular metrics, that we expect no more than two horizons. 
Typically, two distinct horizons are observed when $\mu < \mu_{c}$, where $\mu_{c}$ is a critical ratio for which the two horizons coincide giving us an extremal black hole. Finally when $\mu>\mu_{c}$ we find an horizonless compact object.
Indeed, this is the case for the metric \eqref{trialmetric} as it can be proven for $D\ge4$. The function $f(r)$ is continuous and differentiable $\forall r \ge0$. Asymptotically, $f(r\rightarrow0)=f(r\rightarrow\infty)=1$. Moreover, the positive-definiteness of the second term guarantees $f(r)\le1$. If $f(r)$ only has one extremum in the region $r\in (0,\infty)$ it must be a finite global minimum. It is possible to show that the last requirement is satisfied by \eqref{fR}. 
Then, to conclude our argument on the number of horizons, we 
rewrite the condition \eqref{fR} as
\begin{equation}
 \dfrac{1}{x^{D-3}}\dfrac{\gamma\left(\frac{D-1}{2},\frac{x^2}{4}\right)}{\Gamma\left(\frac{D-1}{2}\right)}=\mu^{-1}.
\end{equation}
The previous discussion implies that the LHS will have only one maximum: increasing $\mu$, the intersection with the line $y(x)=\mu^{-1}$ will provide 0, 1 or 2 horizons. Clearly, more than two horizons are forbidden by the presence of just one extreme. 
The physically different situations, shown in Figure \ref{fR_graph}, are:

\begin{itemize}
\item A dirty gravastar configuration, realized for $0 <R_s^{D-3} <\mu_c\,\ell^{D-3}$, corresponding to a solution with no horizons. Thus, 
\begin{equation}
    M_c=\dfrac{ \mu_c\,\ell^{D-3}}{G_D}\dfrac{\pi^{\frac{D-3}{2} }(D-2)}{8\Gamma\left(\frac{D-1}{2}\right)}
\end{equation}
is a lower bound for the mass of a static black hole given by \eqref{trialmetric};
\item An extremal black hole with one degenerate horizon $r_0=r_+ = r_-$, realized when $R_s^{D-3} =\mu_c\,\ell^{D-3}$; 
\item A two horizons black hole, observed when $R_s^{D-3}>\mu_c\,\ell^{D-3}$, with $\mu_c$ to be determined numerically after fixing $D$. This compact object is a non-extremal black hole with two horizons $r_+ > r_-$.
\end{itemize}
In the classification above we have defined
\begin{equation}
    \mu_c=\max_{x\in\mathbb{R}_{\ge0}} \,\dfrac{1}{x^{D-3}}\dfrac{\gamma\left(\frac{D-1}{2},\frac{x^2}{4}\right)}{\Gamma\left(\frac{D-1}{2}\right)},
\end{equation}
numerically computed to be $\mu_c\sim 3.8$ in four dimensions, and increasing parametrically with $D$.\\
The different classes of solutions discussed above show interesting departures with respect to the Schwarzschild metric when exploring features connected to the horizon structure, such as the asymptotic gravitational redshift and the Hawking temperature.
We focus on $D=4$ for concreteness. The asymptotic gravitational redshift 
\begin{equation}\label{redshif}
    z(r) = \dfrac{\sqrt{g_{tt}(r_{\rm Asympt}\rightarrow\infty)}}{\sqrt{g_{tt}(r)}}-1= e^{\frac{R_S}{2\sqrt{\pi}\ell}e^{-\frac{r^2}{4\ell^2}}} \l[1-\dfrac{2R_S}{r} \dfrac{\gamma(3/2,r^2/4\ell^2)}{\sqrt{\pi}}\r]^{-\frac{1}{2}}  - 1
\end{equation}
differs significantly from a Schwarzschild metric (Figure \ref{redshift_graph}) and, in the horizonless case, is well-defined up to the small $r/\ell$ limit where

\begin{align}
    z(r)&\stackrel{r\rightarrow0}{\rightarrow} \l(e^{\frac{R_S}{2\ell\sqrt{\pi}}}-1\r).
\end{align}
In the case of the black hole solution, we can also compute the Hawking temperature given by
\begin{equation}
    T(r_H,\ell) = \frac{1}{4\pi}\dfrac{1}{\sqrt{|g_{tt}g_{rr}|}}\l. \frac{d|g_{tt}|}{dr}\r|_{r_H}, 
\end{equation}
where $r_H$ denotes the outer horizon. We find
\begin{equation}
    T(r_H,\ell) = \frac{R_s}{ 4\pi^2  \ell^3 r_H^2} e^{-\frac{ R_s e^{-\frac{r_H^2}{4 \ell^2}}}{ 2\ell\sqrt{\pi}}} \left(2 \sqrt{\pi } \ell^3 -r_H^2 R_se^{-\frac{r_H^2}{4 \ell^2}}\right) \gamma \left(\frac{3}{2},\frac{r_H^2}{4 \ell^2}\right).
\end{equation}
The black hole temperature can also be expressed using the defining equation \eqref{fR} for $r_H$:
 \begin{equation}\label{Temperature}
     T(r_H, \ell) = \frac{1}{4\pi r_H}\exp\l[-\frac{r_H}{4\ell}\frac{e^{-\frac{r_H^2}{4\ell^2}}}{\gamma\l( \frac{3}{2},  \frac{r_H^2}{4\ell^2} \r)}\r] \l[1-\frac{r_H^3e^{-\frac{r_H^2}{4\ell^2}}}{4\ell^3\gamma\l( \frac{3}{2},  \frac{r_H^2}{4\ell^2} \r)}\r].
 \end{equation} 
This is only defined and physically sensible for $r_H \geq r_{0}$, the radius of the horizon in the extremal case.
Notice that, when $\ell/r_{H} \to 0$ we recover the usual Schwarzschild black hole temperature
\begin{equation}
T_{\text{Sch}} = \frac{1}{4 \pi r_{H}}\, .
\end{equation}
\begin{figure}[t]
  \centering
  \includegraphics[
  width=0.73\textwidth]{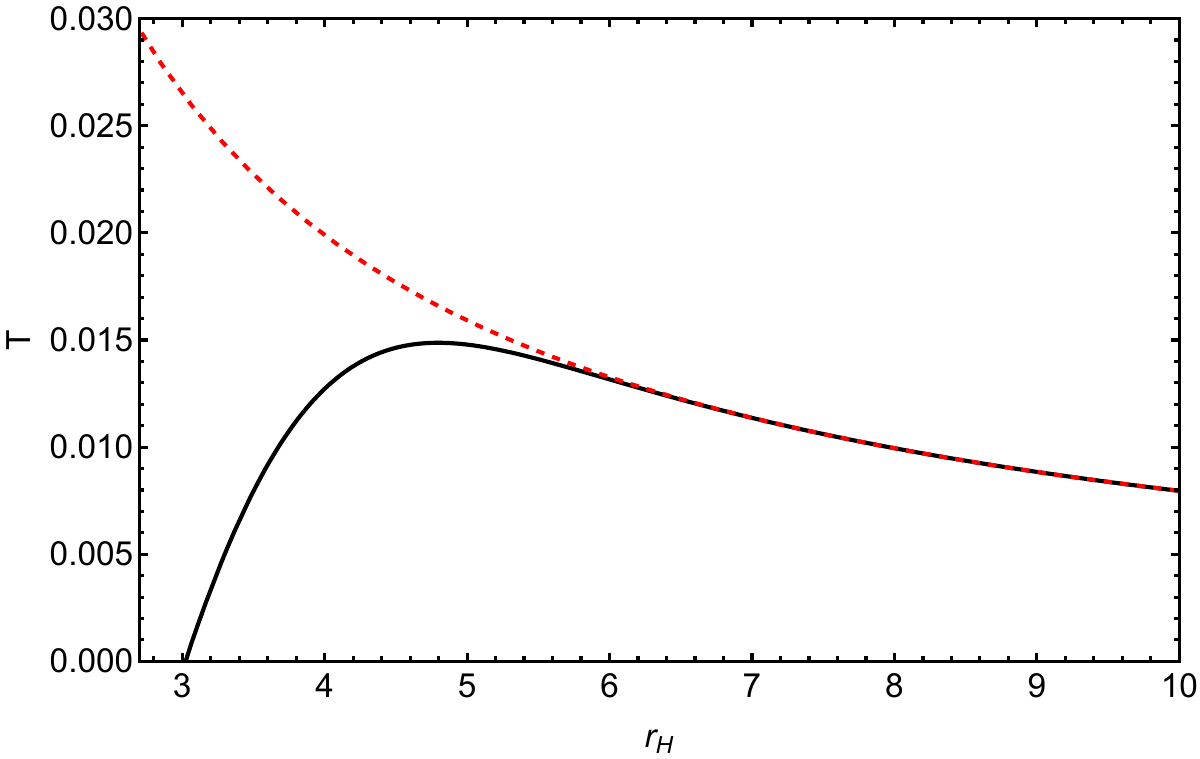}
  \caption{\label{Temperature_graph}
Hawking temperature $T(r_{H})$ of the nonlocal dirty geometry (in black) as a function of its horizon radius $r_H$. In red, the Schwarzschild hyperbola, sharing its large $r_H$ asymptotics. The extremal horizon radius is at $r_0 \approx 3\ell$ and units are chosen so that $\ell =1$.}
\end{figure}A plot of the temperature is shown in Figure \ref{Temperature_graph}.
Although the final state of such regular black holes cannot be rigorously determined without including backreaction effects, semiclassically
a cold remnant has to be expected in agreement with other models sharing the same qualitative and non-monotone behavior of the temperature \cite{Cadoni:2022chn,Bonanno:2006eu}.

\section{Beyond analyticity: strong infrared form factors}
\label{sect:infrared}
A completely different kind of nonlocality, accounting for infrared modifications of Einstein's GR, can be described via non-analytic form factors. In particular, they can effectively describe cosmic acceleration in our Universe \cite{Deser:2007jk} or dynamical dark energy \cite{Maggiore:2014sia}.
Notice that in deriving the leading eikonal for a form factor \eqref{gammaBox}, we never explicitly used the analyticity of $e^{H(-\alpha\Box)}$ and its absence of zeroes in the finite complex plane. Moreover, as already pointed out, the effective modification due to nonlocality manifests into a $e^{-H(\alpha q^2)}$ factor appearing in the propagator \eqref{grav_prop}. Hence, we can compute the effect of a generic $\omega(-{\alpha}\Box)$ by replacing in the eikonal 
\begin{equation}
 e^{-H(\alpha q^2)}\rightarrow\dfrac{1}{1-q^2\omega(\alpha q^2)}.
\end{equation}
Let us start with a form factor \cite{Deser:2007jk,Ferreira:2013tqn} $\omega(\Box)=c{(\Box-\mu^2)^{-1}}$ with $c$ dimensionless and $\mu$ a mass infrared-regulator, 
yielding a leading eikonal (in the probe limit)
\begin{align}
   & 2\delta_0  = \frac{\kappa_D^2M}{\sqrt{E_1^2-m_1^2}}\l[\frac{(D-2)E_1^2-m_1^2}{D-2}\r] \int \frac{d^{D-2}q}{(2\pi)^{D-2}}e^{iqb}\frac{ 1+\mu^2/q^2}{q^2(c+1)+\mu^2}.
\end{align}
Notice how the 1PM deflection angle 
\begin{align}\nonumber
    \Theta^{(1)}=&\chi(D)\Big(\dfrac{R_S}{b}\Big)^{D-3}\Bigg[\dfrac{E_1^2-\frac{m_1^2}{D-2}}{E_1^2-m_1^2}\Bigg]\times\\
    &\l[2^{\frac{D-4}{2}}\Gamma\l(\dfrac{D-2}{2}\r)-c(1+c)^{-\frac{D+2}{4}}(b\mu)^{\frac{D-2}{2}}K_{(D-2)/2}\l(\frac{b\mu}{\sqrt{1+c}}\r)\r]
\end{align}
becomes a rescaled version of Einstein's GR one in the limit $\mu\rightarrow0$. As a check, the case $(D=4,\,m_1=0,\,\mu=0)$ agrees with what found in \cite{Buoninfante:2020qud} for the motion of massless probes in the background of the linearized metric \eqref{smeared_schw}, with
\begin{equation}
    \Phi=-\frac{1}{c+1}\frac{GM}{r}.
\end{equation}
Furthermore, we can try to reconstruct a regular and consistent metric, taking advantage of the eikonal compatibility criterion. 
Solving \eqref{system} for the given form factor and linearizing the Ansatz \eqref{Ansatz} to 1PM, we find
\begin{align} \label{massiveIR1metric} 
  \nonumber ds^2&=-\l[1-\l(\frac{R_s}{r}\r)^{(D-3)}h(\mu r)\r]dt^2 + \l[1+\l(\dfrac{R_s}{r}\r)^{(D-3)}g(\mu r)\r]dr^2 + r^2d\Omega_{D-2},\\
  h(\mu r)&=\dfrac{\Gamma \l(\frac{D-3}{2} \r)-2c(1+c)^{-(D+1)/4}(\mu r/2)^{(D-3)/2}K_{(D-3)/2}\l(\frac{\mu r}{\sqrt{1+c}}\r)}{\Gamma \l(\frac{D-3}{2} \r)},\nonumber\\
  g(\mu r)&=\dfrac{\Gamma \l(\frac{D-1}{2} \r)-2c(1+c)^{-(D+3)/4}(\mu r/2)^{(D-1)/2}K_{(D-1)/2}\l(\frac{\mu r}{\sqrt{1+c}}\r)}{\Gamma \l(\frac{D-1}{2} \r)}.
\end{align}
Again, the limit $\mu\into0$ reproduces a familiar result: $(h(x), g(x))\rightarrow(c+1)^{-1}$ become constants, and the whole metric becomes Schwarzschild with a rescaled normalization. Trivially, the latter is recovered for $c=0$ and independently of $\mu$. 
Though we might be tempted to extend the 1PM metric \eqref{massiveIR1metric} to a nonlinear one in the same fashion as of
Section \ref{guess}, we soon realize that a nonlinear completion of the suggested form still has singularity problems. The singularity at the Killing horizon could be dealt with via higher-order PM corrections in $g_{tt}$, as suggested by \eqref{sigma_gtt}. Nevertheless, looking at the $r\sim0$ behavior of $g_{tt}$ in \eqref{massiveIR1metric}, we realize there is no natural Ansatz to achieve regularity. The singularity problems shared with Schwarzschild suggest instead to consider a different action as a starting point.\\
Let us now  move to a second infrared form factor \cite{Cusin:2015rex}: 
\begin{equation}
    \omega\l(\dfrac{\Box}{\mu^2}\r)=-\dfrac{\mu^2}{\Box^2}
\end{equation}
with $\mu$ acting as an effective mass in the spin-2 component of the graviton propagator. 
Indeed, applying the rule
\begin{equation}
    e^{-H (q^2/\mu^2)}\rightarrow\dfrac{1}{1+\mu^2/q^2},
\end{equation}
the propagator \eqref{grav_prop} takes the form 
\begin{equation}
    G_{\mu\nu;\rho\sigma}(q^2) = -i\frac{1}{q^2+\mu^2-i0}P_{\mu\nu; \rho\sigma}.
\end{equation}
The leading eikonal phase is then 
\begin{equation}
     2\delta_0  = \frac{\kappa_D^2M}{\sqrt{E_1^2-m_1^2}}\l[\frac{(D-2)E_1^2-m_1^2}{D-2}\r] \int \frac{d^{D-2}q}{(2\pi)^{D-2}}e^{iqb} \frac{1}{q^2+\mu^2},
\end{equation}
from which results a 1PM deflection angle of the following form:
\begin{equation}
    \Theta^{(1)}=\chi(D)\l(\dfrac{R_S}{b}\r)^{D-3}\Bigg[\dfrac{E_1^2-\frac{m_1^2}{D-2}}{E_1^2-m_1^2}\Bigg]\,(b\mu)^{\frac{D-2}{2}}\,K_{\frac{D-2}{2}}(b\mu).
\end{equation}
The case $(D=4, m=0)$ recovers the prediction \footnote{To be precise, the leading eikonal phase in this case predicts 
   $ \Theta^{(1)}=\Big(\dfrac{R_S}{b}\Big)\,(2b\mu)\,K_{1}(b\mu).$ This result indeed agrees, setting $a=0$, to Eq. (95) in \cite{Buoninfante:2020qud}, after adding to the $y-$integrand a total derivative $\frac{4GM}{b}\partial_y(\sqrt{1-y^2}e^{-\mu b/y})$ which integrates to 0, and comparing with common integral formulas for modified Bessel functions.
} for a massless probe moving in the background metric  of the form \eqref{smeared_schw} \cite{Buoninfante:2020qud}, but now with a Yukawa potential
\begin{equation}
    \Phi(r)=-\frac{GM}{r}e^{-\mu r},
\end{equation} whereas the Schwarzschild limit is recovered for $\mu b\rightarrow0$, effectively coming back to a massless spin-2 mode.\\
Regarding the reconstruction of a regular and consistent metric, eikonal compatibility is again employed and leads to the linearized 1PM metric
\begin{align} \label{Besselmetric} \nonumber
    ds^2= &-\l[1-2\l(\frac{R_s}{r}\r)^{(D-3)}\frac{(\mu r/2)^{(D-3)/2}K_{(D-3)/2}(\mu r)}{\Gamma \l(\frac{D-3}{2} \r)}\r]dt^2 + \\ &\l[1+2\l(\dfrac{R_s}{r}\r)^{(D-3)}\dfrac{(\mu r/2)^{(D-1)/2}K_{(D-1)/2}(\mu r)}{\Gamma \l(\frac{D-1}{2} \r)}\r]dr^2 + r^2d\Omega_{D-2}.
\end{align}
However, for the same reasons as the $\Box^{-1}$ case, the nonlinear extensions of the form \eqref{Ansatz} are still singular at the origin. 
Absence of a straightforward metric regularization at short-distance is something which might be expected when studying solely infrared form factors, since they are known to play a role mainly in the large-distance physics. 

\section{Conclusions}\label{Conclusion}
In this work we have investigated the leading gravitational eikonal in a general class of nonlocal $D$ dimensional theories of gravity. As one expects, the net effect of nonlocality is to implement a smearing procedure on energy distributions. This was made quantitative both in massless scattering, whereby we have shown the connection with generalized Aichelburg-Sexl geometries, and in the probe limit of massive scattering , which directly allows an interpretation as motion of the probe in a linearized and smeared Schwarzschild background. In both cases, we have shown that the precise shape and extent of the smearing --- and its (weakening) effects on gravitational IR observables --- are controlled by the form factor $e^{H(\alpha\Box)}$ which appears directly at the level of the action.\\
The collisions of two generalized Aichelburg-Sexl shockwaves of the type derived in Section \ref{GAS}, both at impact parameter $b=0$ and at $b\neq0$ have been studied in \cite{Kohlprath_2002}, expanding on the results of Giddings and Eardley \cite{Eardley:2002re}. They provide useful criteria to diagnose the onset of dynamical gravitational collapse, according to whether one can manage (or not) to identify a Marginally Trapped Closed Surface (MTCS) in the geometry just before the collision, which, by causality, is just the linear superposition of the two shocks. It would be interesting, and very natural, to analyze the outcome of the application of these criteria to our analysis. However, at present, in action-based nonlocal theories neither the issue of causality, nor the clear role (if any) played by singularities and singularity theorems, is transparent, so that we couldn't really draw trustworthy results. We may leave it for future works.\\
The deep connection between the eikonal in the probe limit and geometric motion allowed us to put forward a proposal for the reconstruction of the full form of the nonlinear metric generated by a delta source, which is still an open problem in the class of theories under consideration. While there are surely ambiguities in some steps of the reconstruction, our result \eqref{trialmetric}, describing a nonsingular black hole with a dS core, seems consistent and physically motivated, and can be taken as a serious starting point for future investigations. A clear lesson to be drawn from our analysis is the need to go beyond geometries parameterized by a single function, in favour of the richer parametrization of the type
\begin{eqnarray}
    ds^2 = -e^{G(r,\ell, R_s)}F(r,\ell, R_s)dt^2 + F^{-1}(r,\ell,R_s)dr^2 + r^2 d\Omega_{D-2}.
\end{eqnarray}
 The key ingredients in the reconstruction have been the absence of curvature singularities everywhere in spacetime and a particular consistency condition for observables. Regularity of the spacetime at $r=0$ is already granted by the softening of the Newtonian potential, whereas regularity at the would-be horizons has been  enforced choosing a particular form of the metric, in line with the general expectations about the smearing properties of nonlocality. Inverting the reasoning, nonlocality can offer a compelling, first principles interpretation of a wide family of regular spacetimes.\\
Much of the eikonal physics in nonlocal field theories of gravity remains uncharted territory. A natural extension of our tree level results would be to analyze higher PM orders, with an explicit computation of subleading terms in higher loop amplitudes. Already at one loop and in the probe limit, this would allow to test if \eqref{trialmetric} correctly predicts also the 2PM deflection angle. If positive, the test would strongly support our proposal. If negative, we expect to be able to refine the Ansatz employing the same consistency conditions. Another instructive computation would be the classical limit of the $2 \rightarrow3$ amplitude involving, besides two massive scalars, an additional graviton emission. This is known to be directly linked to the leading order PM waveform \cite{DiVecchia:2021bdo,Goldberger:2016iau,Luna:2017dtq} and could thus lead to clear and potentially observable signatures of nonlocal physics.\\
A third natural and interesting development, though technically somewhat difficult to implement, could consist in employing the very same procedure, i.e. conjecture a general form of the metric for the regular spacetime and then impose  \emph{eikonal compatibility}, to build rotating regular black holes via scattering amplitudes, along the lines of \cite{Bianchi:2023lrg}. An intermediate check would again be provided by comparison with deflection angles evaluated in \cite{Buoninfante:2020qud} in the slow rotation limit. First steps in this direction could be made employing the Newman-Janis algorithm, see e.g. \cite{Bambi:2013ufa}, to static regular solutions as \eqref{trialmetric}.
\\
Apart from assuming its decoupling from the Planck scale --- necessary for the implementation of a classical limit --- we have willingly left unspecified the nonlocality scale and its nature. A reason for doing so is that ghost-free nonlocal theories of gravity arise as the result of a careful procedure in which, compared to GR, new degrees of freedom are forbidden (except for a single scalar field useful to drive inflation \cite{Briscese:2013lna}) and UV super-renormalizability or finiteness are achieved \cite{Krasnikov:1987yj,Kuzmin:1989sp,Tomboulis:2015esa,Modesto:2011kw,Modesto:2014lga,Modesto:2017sdr,Calcagni:2023goc,Briscese:2018oyx,Briscese:2021mob}. In them, the nonlocality scale $\ell$ appears as an unspecified parameter and can be tuned at will to explore phenomenological or conceptual implications of nonlocality. For example, a superplanckian $\ell$ in the spacetime metric may have observable signatures  in black hole imaging by the EHT collaboration \cite{EventHorizonTelescope:2019dse}, in orbital dynamics \cite{Cadoni:2022vsn}, and gravitational wave signals soon to be detected by third generation interferometers like Einstein Telescope \cite{ET:2025xjr}; this last feature would be partially shared by a Planckian $\ell$ whose effects could be magnified by gravitational interactions of a large number of particles \cite{Buoninfante:2019teo,Buoninfante:2018gce,Buoninfante:2019swn}. 
Though we saw hints of qualitative similarities with other nonlocal approaches, such as string theory or  noncommutative geometry, a microscopic understanding of the precise form of the action of the theories studied in this paper is missing. A practical stance in regard to this matter is to view nonlocality as just a fundamental feature of spacetime --- and gravitational interactions --- at a certain scale, and work out the consequences.

\acknowledgments
We are grateful to Andrea P. Sanna for his numerous suggestions for improving our work.

\newpage
\newpage
\appendix
\section{Conventions and kinematics}
\label{app:A}
In the following we spell out for convenience of the reader the main conventions that we employ. 
Throughout the paper, unless if they appear explicitly, we adopt natural units $\hbar = c =k_B = 1 $.
By convention, all external momentum vectors will be regarded as outgoing, so that $-p_{1}$ and $-p_{2}$ will represent the physical momenta for the two incoming particles. We work with the mostly-plus signature for the metric,
\begin{equation}
    \eta_{\mu\nu} = \text{diag} (-1,\,1,...,\, 1)\, \quad \mu,\nu = 0,\, 1,...,\, D-1.
\end{equation}
We label the incoming particles by $(1,2)$ and the outgoing ones by $(3,4)$, in such a way that
\begin{equation}
    p_{1}^{2} = p_{4}^{2} =  - m_{1}^{2}\, \quad p_{2}^{2} = p_{3}^{2} = - m_{2}^{2}.
\end{equation}
The external momenta satisfy the momentum conservation condition
\begin{equation}
    p_{1} + p_{2} + p_{3} + p_{4} = 0
\end{equation}
and we define the usual Mandelstam variables via
\begin{equation}
    s = -(p_{1} + p_{2})^{2}\, \quad t = - (p_{1} + p_{4})^{2}\, \quad u = - (p_{1} + p_{3})^{2}.
\end{equation}
The total energy in the center-of-mass frame $E$ is given by $E = \sqrt{s}$, while the momentum transfer
\begin{equation}
    q = p_{1} + p_{4} = - p_{2} - p_{3}
\end{equation}
 is related to the Mandelstam variable $t = - q^{2}$. Finally, $u$ can be written in terms of $s$ and $t$ using momentum conservation and the mass-shell conditions,
\begin{equation}
    s + t + u = 2 (m_{1}^{2} + m_{2}^{2}).
\end{equation}
Another ubiquitous variable is the relative Lorentz gamma factor
\begin{equation}
    \sigma \equiv \frac{1}{2} \l(\frac{s-m_1^2-m_2^2}{m_1 m_2}\r),
\end{equation}
In the center of mass frame, the initial momenta can be taken as
\begin{equation}
    -p_1 = (E_1,\vec{p}), \qquad -p_2= (E_2,-\vec{p}),
\end{equation}
and, defining $p \equiv \l|\:\vec{p}\:\r|$, the following relations hold
\begin{align}
    Ep &=  m_1 m_2\sqrt{\sigma^2-1},\\
    E &= E_1+E_2 = \sqrt{m_1^2+m_2^2+2m_1m_2\sigma}.
\end{align}
We will denote every amplitude, stripped down of the overall momentum conserving factor, as $\Amplitude$. Writing the S-matrix as
\begin{equation}
    S= 1+iT
\end{equation}
we have
\begin{equation}
    \langle p_4,p_3 |T|-p_2,-p_1\rangle = (2\pi)^D \delta^{(D)}(p_1+p_2+p_3+p_4)\mathcal{A}(s,t).
\end{equation}
We can also include the disconnected piece in a normalized S-matrix definition
\begin{equation}
    \mathcal{S}(p_1,p_2;q) = (2\pi)^{D} \delta^{D}(q) + 2\pi\delta(2p_1\cdot q- q^2)2\pi\delta(2p_2\cdot q+ q^2)i\mathcal{A}(s,-q^2).
\end{equation}
Finally, the eikonal phase is obtained from the Fourier transform of the formula above
\begin{equation}
    \tilde{\mathcal{S}} (s,b) = 1+\int \frac{d^{D-2}q}{(2\pi)^{D-2}}e^{ibq} \frac{i\mathcal{A}(s,-q^2)}{4Ep} \quad+\:\:\:... \quad\approx e^{2i\delta(s,b)},
\end{equation}
modulo quantum corrections which spoil the exponentiation but are not relevant for the analysis in the paper.\\
Finally, we shall make use of the following definition for the $D$ dimensional Schwarzschild radius
\begin{equation}\label{Rs}
    R_s^{D-3} = \frac{8G_DM \Gamma \l( \frac{D-1}{2} \r)}{\pi^{\frac{D-3}{2}}(D-2)}.
\end{equation}
\section{Loops and exponentiation}

\label{app:B}
In this appendix, following  closely \cite{DiVecchia:2023frv,Kabat:1992tb,Akhoury:2013yua}, we provide a check for the (massless) eikonal exponentiation mechanism.
The dominant diagrams at each loop order will still be ladders and crossed ladders, since it follows basically from the structure of the vertices and of the energetic scalar propagators, both unchanged with respect to the local case. The leading term in the amplitude for the exchange of $n$ gravitons, at loop order $L=n-1$ will take the form
\begin{equation}
    i\mathcal{A}_{n-1}(s,t) = (-k_Ds^2)^n \int \prod_{j=1}^n\l[\frac{d^D\ell_j}{(2\pi)^D}G(\ell_j)e^{-H(\alpha \ell^2_j)}\r](2\pi)^D\delta^{(D)}(q-\ell)I^{(n)},
    \end{equation}
where $\ell= \sum_{j=1}^n \ell_j $ (not to be confused with the nonlocality scale), the factor $(-k_Ds^2)^n$ comes from the high energy limit of the $2n$ vertices connected pairwise, one upper and one lower, by a graviton propagator, whose scalar part we denote by $G(\ell_j)e^{-H(\alpha \ell_j^2)}$ and $I^{(n)}$, is the structure resulting from the product of scalar propagators coming from the two energetic lines.\\
$I^{(n)}$ must contain a sum over all possible ways of attaching the rungs, encoded in a double permutation sum over the attached momenta in each line, weighted by a $1/n!$ averaging factor which ensures that  diagrams differing only by a relabeling of the internal lines are counted only once. Explicitly
\begin{align}
    I^{(n)} = \frac{1}{n!} \sum_{\sigma \in S_n}G(p_1-\ell_{\sigma_1})G(p_1- \ell_{\sigma_1}-\ell_{\sigma_2})...G(p_1-\ell_{\sigma_1}-...-\ell_{\sigma_{n-1}})\\
    \nonumber
    \sum_{\sigma' \in S_n}G(p_2+\ell_{\sigma'_1})G(p_2+ \ell_{\sigma'_1}+\ell_{\sigma'_2})...G(p_2+\ell_{\sigma'_1}+...+\ell_{\sigma'_{n-1}}).
\end{align}
For the scalars, we can use the high energy approximation in the propagators
\begin{equation}
    G(p_1-\ell_k) =  \frac{-i}{(p_1-\ell_k)^2 -i0} \approx\frac{-i}{-2p_1\cdot \ell_k -i0}.
\end{equation}
The next step is a formal rewriting of $\delta^{(D)}(q-\ell)$ as
\begin{equation}
    \delta^{(D)}(q-\ell) \approx 2s \delta(2p_1\cdot \ell)\delta(2 p_2 \cdot \ell)\delta^{(D-2)}(q_\perp -\ell_\perp).
\end{equation}
To see this, write a Sudakov decomposition
\begin{equation}
    \ell=\alpha p_1 + \beta p_2 +\ell_\perp.
\end{equation}
Since $s=-2p_1 \cdot p_2$ and $\ell_\perp$ is defined to be transverse to both $p_1$ and $p_2$, we find
\begin{align}
2\ell\cdot p_2 = -s\alpha, \\ 2\ell\cdot p_1 = -s\beta.
\end{align}
The Jacobian can be extracted from 
\begin{equation}
    (d\ell)^2 =-sd\alpha d\beta +d\ell_\perp^2
\end{equation}
as
\begin{equation}
    \sqrt{-\det\gamma_{\ell}} = \frac{s}{2}.
\end{equation}
The delta function transforms with the inverse Jacobian
\begin{align}
    \delta^{(D)}(q-\ell) = \frac{2}{s}\delta(\alpha_q-\alpha)\delta(\beta_q-\beta)\delta^{(D-2)}(q_\perp-\ell_\perp)=\\
    \nonumber
    \frac{2}{s}\delta(-2p_2\cdot q/s+2p_2\cdot \ell/s)\delta(-2p_1\cdot q/s +2p_1\cdot \ell/s)\delta^{(D-2)}(q_\perp-\ell_\perp)\\\nonumber
    \approx  2s \delta(2p_1\cdot \ell)\delta(2 p_2 \cdot \ell)\delta^{(D-2)}(q_\perp -\ell_\perp).
\end{align}
The two delta functions can then be combined with the permutation sums to give the  distributional identity
\begin{equation}
    \delta(\omega_1+...+\omega_n) f(\omega_1-i0, \omega_2-i0,...\omega_n -i0) = (2i\pi)^{(n-1)}\delta(\omega_1) \delta(\omega_2)...\delta(\omega_n)
\end{equation}
where
\begin{equation}
    f(a_1,...,a_n) = \sum_{\sigma \in S_n} \frac{1}{a_{\sigma_1}} \frac{1}{a_{\sigma_1} +a_{\sigma_2}}...\frac{1}{a_{\sigma_1}+...+a_{\sigma_{n-1}}}
\end{equation}
and for the two lines we have $a_{\sigma_j} = 2p_i \cdot \ell_{\sigma_j}, i=1,2$.
The outcome of these manipulations is to enforce a transversality condition also on the loop momenta, yielding an almost factorized form
\begin{align}\nonumber
    i\frac{\mathcal{A}_{n-1}(s,t)}{2s} =& \frac{(-k_Ds^2)^n}{n!}\int\prod_{j=1}^{n} \l[ \frac{d^D\ell_j}{(2\pi)^{D-2}}e^{-H(\alpha \ell_j^2)}G(\ell_j) \delta(2p_1\cdot \ell_j)\delta(2p_2\cdot \ell_j)\r] \\
    &(2\pi)^{D-2}\delta^{(D-2)}(q_\perp -\ell_\perp).
\end{align}
To completely factorize the result, we take
\begin{equation}
    \tilde{\mathcal{A}}(s,b) = \int \frac{d^{D-2}q}{(2\pi)^{D-2}}e^{iq_\perp \cdot b} \frac{\mathcal{A}(s,-q_\perp^2)}{2s}.
\end{equation}
This gives
\begin{equation}
    i\tilde{\mathcal{A}}_{n-1}(s,b) = \frac{(-k_Ds^2)^n}{n!} \l[\int \frac{d^{D}\ell}{(2\pi)^{D-2}}G(\ell) \delta(2 p_1 \cdot \ell) \delta(2 p_2 \cdot \ell)e^{-H(\alpha \ell^2)}\r]^n,
\end{equation}
So that
\begin{equation}
    1+ i \sum_{n=1}^{\infty} \tilde{\mathcal{A}}_{n-1}(s,b) \approx e^{2i\delta_0 (s,b)}
\end{equation}
and 
\begin{equation}
    2i\delta_0 = i \tilde{\mathcal{A}_0}(s,b)
\end{equation}
is the leading eikonal.\\
As anticipated, the form factor enters the calculation entirely as a spectator, at least for the leading result.
\section{Geodesic motion}
\label{app:C}
We wish to study geodesic motion of a spinless probe in the background described by a metric of the form \eqref{Ansatz}. The generic starting point is the action
\begin{equation}
    S_p = \frac{1}{2}\int d\tau \l[ \frac{1}{e(\tau)} g_{\mu\nu} \dot{x}^\mu \dot{x}^{\nu} -m^2_pe(\tau)\r].
\end{equation}
Owing to time independence and spherical symmetry, there are two conserved quantities, the energy $E_p$ of the probe and its angular momentum $J$
\begin{equation}
    eE_p = \l|g_{tt}\r| \dot{\tau}, \qquad eJ = r^2 \bar{g}_{\phi\phi} \dot{\phi}.
\end{equation}
Where we parameterized $g_{\phi\phi} \equiv r^2\bar g_{\phi\phi}$, and $\phi$ is the angle in the scattering plane singled out by spherical symmetry. \\
The largest root of $\dot{r}$ defines the inversion point $r_*$ of the trajectory, i.e.
\begin{equation}
    \dot{r} |_{r_*}=0,
\end{equation}
or, more explicitly, varying the action and using the definition of the affine parameter
\begin{equation} \label{inversion}
    \l[\frac{E_p^2}{g_{rr} \l|g_{tt}\r|} - \frac{1}{g_{rr}} \l(\frac{J^2}{r^2\bar g_{\phi\phi}} +m_p^2\r)\r]_{r_*}=0.
\end{equation}
As the trajectory is symmetric about the inversion point, the total scattering angle is twice the accumulated $\frac{d\phi}{dr} dr$ from $r_*$ to infinity (minus $\pi$)
\begin{equation}\label{theta}
    \Theta = 2J \int_{r_*}^{\infty} \frac{dr}{r^2} \l[\dfrac{\bar{g}_{\phi\phi}^2E_p^2}{g_{rr} \l|g_{tt}\r|} - \dfrac{\bar g_{\phi\phi}}{g_{rr}} \l(\dfrac{J^2}{r^2} + \bar g_{\phi\phi}m_p^2\r)\r]^{-1/2} -\pi.
\end{equation}
We are interested in the 1PM expansion of this quantity, valid when $r_*$ is small compared with $R_s$. From \eqref{inversion} and the metric components in \eqref{Ansatz} we find that, up to higher order corrections, we can write
\begin{equation}
    r_* \approx r_0 \l[1+ \bar\delta \l(\frac{R_s}{r_0}\r)^{D-3}\r],
\end{equation}
with
\begin{equation}
    \bar{\delta}= - \frac{g(r_0/\ell)E_p^2}{2(E_p^2-m_p^2)} + \frac{E_p^2}{2(E_p^2-m_p^2)} (g(r_0/\ell)- h(r_0/\ell)) = - \frac{E_p^2}{2(E_p^2-m_p^2)}h(r_0/\ell),
\end{equation}
and 
\begin{equation}
    r_0 = \frac{J}{\sqrt{E_p^2-m_p^2}}
\end{equation}
can be identified, to this order, with the impact parameter $b$ appearing on the eikonal side. We can now proceed by expanding the integrand \eqref{theta} to order $(R_s/r_0)^{D-3}$, after changing variables to $u = \frac{r_*}{r}$
\begin{align}\nonumber
    &\Theta +\pi \approx \frac{2J}{r_* \sqrt{E_p^2-m_p^2}}\int \frac{du}{\sqrt{1-u^2}} \bigg\{1+ \frac{1}{2}\l(\frac{R_s}{r_0}\r)^{D-3} \\
    & \bigg[ \frac{E_p^2u^{D-3}[g\l(r_0/\ell u)- h(r_0/\ell u)\r] -2(E_p^2-m_p^2)u^2\bar\delta -u^{D-3}g(r_0/\ell u)\l[ (E_p^2-m_p^2)u^2+m_p^2\r] }{(1-u^2)(E_p^2-m_p^2)}\bigg]\bigg\}.
\end{align}
Further expanding the prefactor, combining the terms and setting $r_0 =b$, we finally get
\begin{align} \label{geothetader}\nonumber
    \Theta^{(1)} = &\l(\frac{R_s}{b}\r)^{D-3}\frac{1}{E_p^2-m_p^2} \int_0^1 \frac{du\,u^{D-3}}{(1-u^2)^{3/2}}\bigg\{E_p^2\l[g\l(\frac{x}{u}\r) -h\l(\frac{x}{u}\r) \r] \\
    &-g\l(\frac{x}{u}\r) \l[ (E_p^2-m_p^2)u^2 +m_p^2\r] \bigg\}, \:\:\: \text{with} \:\:\: x\equiv b/\ell.
\end{align}
We have kept exact the dependence in $(r/\ell)$, which is consistent provided the functional form of $g(r/\ell)$ and $h(r/\ell)$ does not interfere with the $R_s/r$ expansion. This is the case, for example, with functions ranging in $[0,1]$, for every value of $r/\ell$ (even when $\ell \sim R_s$), such as the majority of those we extracted in the main body of the paper. We expect it be satisfied, in the general case, also when there is hierarchy of scales, such as $\ell \ll R_s$. In this case, one can keep the two expansions distinct and study systematically the corrections.

\bibliographystyle{JHEP}
\bibliography{biblio}
\end{document}